%% file: FinalVersion_OTA_Journal.tex
\input{header.tex}

\usepackage{soul}

\title{{Over-the-Air Fronthaul Signaling for\\ Uplink Cell-Free Massive MIMO Systems}}

\author{
\medskip
\normalsize
Zakir Hussain Shaik,
{\em Member, IEEE}, Sai Subramanyam Thoota, {\em Member, IEEE}, \\Emil Bj\"{o}rnson, {\em Fellow, IEEE}, and Erik G. Larsson, {\em Fellow, IEEE}
\normalsize
\thanks{Preliminary results of this paper were presented in IEEE ICC 2024 \cite{zhshaik_icc2024}.} 
\thanks{Z. H. Shaik and S. S. Thoota were with the Department of Electrical Engineering (ISY), Linköping University, SE-58183 Linköping, Sweden, at the time of manuscript submission.
E. G. Larsson is with the Department of Electrical Engineering (ISY), Linköping University, SE-58183 Linköping, Sweden. They were supported in part by KAW foundation and ELLIIT.}
\thanks{E. Bj\"ornson is with the Department of Computer Science, KTH Royal Institute of Technology, SE-10044, Stockholm, Sweden. He was supported by the Grant 2019-05068 from the Swedish Research Council.}
\thanks{Emails: zakir.hussain.shaik@liu.se, sai.subramanyam.thoota@liu.se, emilbjo@kth.se, erik.g.larsson@liu.se.}}
 
\allowdisplaybreaks
\begin{document}

\maketitle
\date{}
\begin{abstract}
We propose a novel resource-efficient \gls{ota} computation framework to address the huge fronthaul computational and control overhead requirements in cell-free massive \gls{mimo} networks. 
We {show} that the global sufficient statistics to decode the data symbols can be computed \gls{ota} using the locally available information {at the \glspl{ap}}. We provide the essential signal processing aspects at the \glspl{ap} and the \gls{cpu} to facilitate the \gls{ota} computation of sufficient statistics. The proposed framework scales effectively with an increase in the number of \glspl{ap}. {We also make a comprehensive study of the benefits of an \gls{ota} framework compared to a conventional digital fronthaul in terms of the overhead associated in transferring the sufficient statistics from the \glspl{ap} to the \gls{cpu}.} To evaluate the performance of the \gls{ota} framework, we give closed-form expressions for the \gls{mse} of the estimators of sufficient statistics and 
the overall data estimator. Furthermore, we assess the \gls{ser} {and} \gls{ber} of the \glspl{ue} data to demonstrate the efficacy of our method, and benchmark them against the state-of-the-art wired fronthaul networks.

\end{abstract}


\section{Introduction}
Cell-free massive \acrfull{mimo} is envisioned as a next-generation wireless technology with the potential to significantly enhance the \gls{qos} of the \acrfull{ue} by leveraging macro-diversity, thereby providing high \gls{se}~{\cite{demir2021foundations}}. However, to meet the demands of future applications, we need to deploy a large number of \acrfullpl{ap}, which translates to an increase in the number of fronthaul interconnections from them to the \acrfull{cpu}. 
Typically, in \gls{ul} cell-free \gls{mimo} systems, high-speed wired fronthaul links are used to transfer the received signals and the \glspl{ue}' \acrfull{csi} from the \glspl{ap} to the \gls{cpu} for centralized data decoding. Therefore, as the number of \glspl{ap} increases, the number of fronthaul connections may become unscalable, which necessitates alternate solutions. Moreover, the wired fronthaul links may restrict the placement of \glspl{ap} due to geographical limitations leading to a degraded system performance.


In order to handle the fronthaul challenges in large cell-free massive \gls{mimo} networks, we propose an \acrfull{ota} framework that is not only scalable with {an} increase in the number of \glspl{ap} but also has minimal performance loss 
{compared to a wired fronthaul}. The superposition properties of electromagnetic waves is the underlying principle governing the \gls{ota} technique. The primary purpose of the \gls{ota} computation is to estimate a function, $f(x_1,\ldots,x_L)\rightarrow \mathbb{C}$ of the signals transmitted by $L$ different nodes, $x_l,\ l \in \{1,\ldots,L\}$. The most common functions that are relevant or studied for \gls{ota} computation in the literature are so-called nomographic functions. A few examples of such functions 
are sum, geometric mean, maximum and minimum functions~\cite{csahin2023survey}. However, to implement \gls{ota} successfully over distributed networks, appropriate pre-and-post processing techniques are needed. 

\colr{Put succinctly, the advantages of an over-the-air fronthaul are first, deployment flexibility: for example,
\glspl{ap} can be easily moved and new \glspl{ap} can be easily installed; and second, cost:
the wireless-fronthaul architecture requires no deployment of high-capacity data link cables to the access points.}
In this paper, we propose  an \gls{ota} framework to address the multiple challenges that arise during the fronthaul signalling in cell-free massive \gls{mimo} networks.

\subsection{Related Work}
%
\Gls{ota} computation has garnered a lot of traction in the literature recently~\cite{atzeni2020distributed,han2023optimized,jing2023transceiver,jiang2021joint,zhai2021hybrid,shao2022bayesian,razavikia2023channelcomp,jha2022fundamental,chen2024over}. 
In \cite{atzeni2020distributed}, the authors designed a distributed \gls{dl} precoder using \gls{ota} computation mechanism while ensuring that the system does not get impacted {by} scalability issues as the the number of \glspl{ap} increases. In \cite{han2023optimized}, a cell-free massive \gls{mimo} network is considered with single-antenna \glspl{ap}, where each \gls{ue} computes a local summed signal and an \gls{ota} computation {is used} to aggregate them. The \glspl{ap} then assists in forwarding the aggregated data back to the \glspl{ue} for a final processing, optimizing the transmit power and receive filters to minimize computation errors. In \cite{jing2023transceiver}, a two-phase hybrid beamforming algorithm is proposed to optimize the transmit and receive beamformers to minimize the \gls{mse} of the sum of the signals with a reduced algorithm execution time. In \cite{jiang2021joint}, the authors proposed an alternating optimization based solution to jointly optimize {the} beamforming matrices at the \glspl{ue}, relays, and the \gls{ap} to estimate the sum of the \gls{ul} signals transmitted \gls{ota}. 
In \cite{zhai2021hybrid}, the authors enhance the computational accuracy of \gls{ota} in \gls{iot} networks by utilizing hybrid beamforming in massive \gls{mimo} systems. They jointly optimize the transmit digital beamforming at the \glspl{ue} and the receive hybrid beamforming at the \gls{ap} to minimize the \gls{mse} of the summed signals. In \cite{shao2022bayesian}, the authors proposed a Bayesian approach for \gls{ota} computation, wherein each edge device transmits statistical information to the fusion center. This information is used to develop Bayesian estimators that are robust to noise and misalignments compared to the traditional ML estimators. In \cite{razavikia2023channelcomp}, the authors developed a method that enables a digital modulation for \gls{ota} by leveraging the benefits of digital communications, such as error correction and synchronization, making it compatible with existing digital systems. However, there are a few limitations on the type of modulation that can be used and also the type of function that can be computed \gls{ota}. In \cite{jha2022fundamental}, the authors study the fundamental limits of analog \gls{ota} computation and analyze the trade-offs between the analog and digital schemes for different \gls{snr} operating regimes. To the best of our knowledge, none of the existing literature develop a framework to compute the sufficient statistics \gls{ota} to decode the information of the \glspl{ue} directly and also provide theoretical performance results to analyze the same in \gls{ul} cell-free massive \gls{mimo} systems. In \cite{chen2024over}, the authors proposed designs for transmit coefficients and receive combining to facilitate \gls{ota} computation in cell-free massive \gls{mimo} with varying levels of cooperation among \glspl{ap}. They demonstrate that cell-free massive \gls{mimo} outperforms conventional cellular \gls{mimo} by reducing the mean squared error, particularly when devices operate with limited power budgets.

\subsection{Motivation}
\colr{One of the typical assumptions in cell-free massive \gls{mimo} systems is that the fronthaul links between the \gls{cpu} and \glspl{ap} are wired~\cite{demir2021foundations}. However, the deployment of these links may be costly. A potential solution is to use wireless communications between the \glspl{ap} and the \gls{cpu}. To centrally decode the \glspl{ue}' data, all the \glspl{ap} need to transmit their received signals and \gls{csi} to the \gls{cpu}. If orthogonal access is used then the number of required resources increases with the number of \glspl{ap}, which also causes scalability issues. This motivates the need for alternative solutions.}

Our aim is to convey all the sufficient information from the \glspl{ap} to the \gls{cpu} to decode the \glspl{ue}' data in a resource-efficient manner still being scalable with increase in the number of \glspl{ap}. This solution also avoids storing data from all the \glspl{ap} separately at the \gls{cpu}. \bluez{Another important point is to reduce the total number of transmissions from the \glspl{ap} such that only the sufficient information is available at the \gls{cpu} to decode the \glspl{ue}' data.} \colr{Incorporating signal processing at \glspl{ap} helps distribute the computational load, reducing the complexity at the \gls{cpu} as the number of \glspl{ap} increases. Moreover, if local processing is properly employed then, the fronthaul load does not scale with the number of \glspl{ap} \cite{shaik2021distributed} as opposed to the case with centralized processing. Furthermore, the benefits of incorporating signal processing at the \glspl{ap} have been extensively studied in the cell-free massive \gls{mimo} literature \cite{demir2021foundations}. Distributed signal processing can improve reliability and enhance network robustness, for instance in the event of \gls{cpu} hardware failure.}

To address all the aforementioned points, we propose a resource-efficient \gls{ota} computation framework to obtain the sufficient statistics transmitted from the \glspl{ap} to the \gls{cpu} in order to decode the \gls{ul} data. Our approach needs only pilot signal broadcast from the \gls{cpu} and all the \glspl{ap} acquire their local fronthaul \gls{csi}.\footnote{\bluez{We consider perfect synchronization between the \glspl{ap} and \gls{cpu} to expose our \gls{ota} framework, and refer the readers to \cite{larsson2024massive,Unnikrishnan_TWC_2024} for detailed studies of synchronization aspects in distributed \gls{mimo} systems.}} Note that the fronthaul \gls{ul} and \gls{dl} links are reciprocal.

\colr{\gls{ota} aggregation has been primarily studied in federated learning, where   gradient updates are
transmitted from edge nodes \cite{gafni2022federated}. In contrast, in cell-free massive MIMO, the  challenge is to identify the data that can be efficiently aggregated from \glspl{ap} to the \gls{cpu}: the answer, we show in this paper, is the \emph{sufficient statistics for the data decoding} (i.e., the Gramians and the matched filter outputs). 
Additionally, since \gls{ota} relies on analog communication, its benefits over digital transmission must be demonstrated. 
In particular, with OTA, appropriate MIMO precoders and decoders, and power control policies, must be designed for the AP-CPU links. This paper addresses these challenges.}

\subsection{Summary of Contributions}
We summarize the main contributions of the paper below:


\begin{enumerate}
    \item We propose a novel scalable and resource-efficient framework to compute the sufficient statistics \gls{ota} to decode the \gls{ul} data in cell-free massive \gls{mimo} systems. 
    \item We develop the transmit precoding and power control policy of the \glspl{ap} using only the local \gls{csi} of their fronthaul links to enable the \gls{ota} computation of the sufficient statistics for \gls{ul} data decoding. 
    \item We provide closed\bluez{-}form expressions for the first and second\bluez{-}order moments required for the Bayesian estimation of the sufficient statistics and also for the power scaling to satisfy the average transmit power constraint. We also present the closed-form expression for the \gls{nmse} of the estimators of the sufficient statistics. 
    Further, we provide a closed-form expression for the \gls{mse} of the \glspl{ue}' data estimator.
    \item We present expressions of the \glspl{ue}' achievable rates for the proposed \gls{ota} framework. We compare them with the achievable rates obtained using a wired fronthaul and demonstrate through numerical analysis that they closely match each other.
    \item We extend our proposed framework to a scenario involving imperfect \gls{csi} and explain how the performance metrics developed with perfect \gls{csi}, such as the \gls{mse} of the sufficient statistics and the \glspl{ue} data, can be adapted to situations with imperfect \gls{csi}.
    \item We compare the proposed \gls{ota} framework to a digital fronthaul scheme, offering a comprehensive analysis and discuss the trade-offs involved. We provide several insights through numerical evaluations.
\end{enumerate}

\section{Cell-free Massive \Acrshort{mimo} and\\ Over-the-Air Computation Framework}\label{sec:SystemModel}
\begin{figure}
\centering
\begin{subfigure}{0.45\textwidth}
    \includegraphics[width=\textwidth]{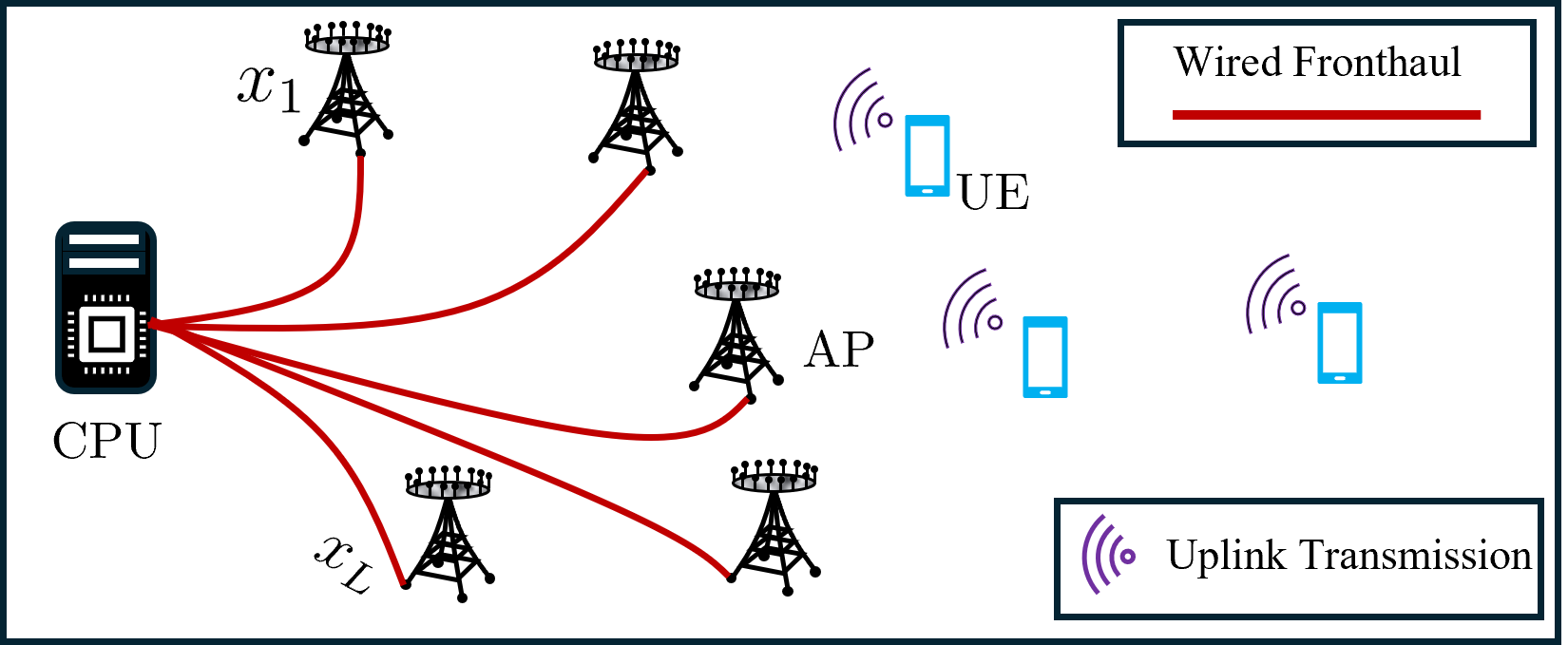}
    \caption{Wired fronthaul to communicate between the \glspl{ap} and the \gls{cpu}.}
    \label{fig:wiredsysmodel}
\end{subfigure}
\hfill
\begin{subfigure}{0.45\textwidth}
    \includegraphics[width=\textwidth]{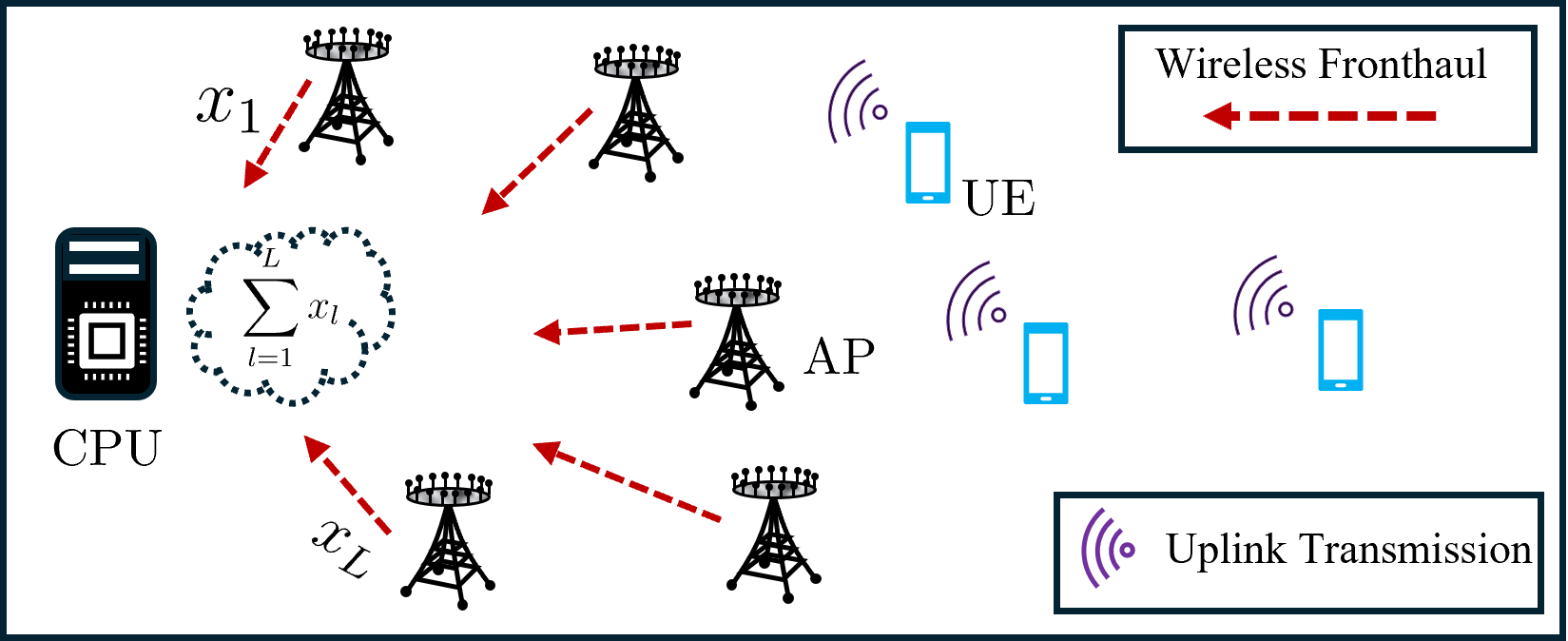}
    \caption{Wireless fronthaul to communicate between the \glspl{ap} and the \gls{cpu}.}
    \label{fig:wirelessysmodel}
\end{subfigure}
        
\caption{Illustration of a cell-free massive \gls{mimo} network.}
\label{fig:sysModel2}
\end{figure}

We consider an \gls{ul} cell-free massive \gls{mimo} system with $K$ single-antenna \glspl{ue} transmitting data to $L$ \glspl{ap} with $N$ receive antennas each. The \glspl{ap} communicate to a central \gls{cpu} equipped with $M$ receive antennas via a wireless fronthaul in a separate frequency band using additional dedicated \gls{rf} chains. We show an illustration of a cell-free massive \gls{mimo} system setup with wired and wireless fronthauls in Fig.~\ref{fig:wiredsysmodel} and Fig.~\ref{fig:wirelessysmodel}, respectively.\footnote{Note that the notation provided in Fig.~\ref{fig:sysModel2} is only for illustration. More details about the \gls{ota} computation framework will be provided subsequently.} We assume that the \glspl{ap} transmit to the \gls{cpu} using $N$ antennas each. We adopt a quasi-static block fading model for all the wireless channels in the network. Note that the number of receive antennas to communicate from the \glspl{ue} to the \glspl{ap} is kept equal to the number of antennas to communicate from the \glspl{ap} to the \gls{cpu} to {keep the} notation simple. However, our proposed approach holds even otherwise with different antenna numbers. 
It can also be generalized when the number of receive antennas at each \gls{ap} is different from each other. We denote the channel between \gls{ue} $k$ and \gls{ap} $l$ by $\mathbf{H}_{l}\in \complexm{N}{K}$ and the channel between the \gls{cpu} and \gls{ap} $l$ by $\mathbf{G}_l \in \complexm{N}{M}$.

We represent the received signal at the $l$-th \gls{ap} during the \gls{ul} data transmission phase from the $K$ \glspl{ue} as
\begin{equation}\label{eqn:recSigAPs1}
	\mathbf{y}_l = \sqrt{p_{\rm ul}}\mathbf{H}_l\mathbf{s} + \mathbf{n}_l \in \complexm{N}{1},
\end{equation}
where $p_{\rm ul}$ is the transmit power of each \gls{ue}, $\chanMat{l} = \left[\chan{1l},\chan{2l},\ldots,\chan{Kl}\right]\in \complexm{N}{K}$ is the channel matrix between {the} $l$-th \gls{ap} and the \glspl{ue}, whose columns are mutually independent and further {the} $k$th column denoted by $\chan{kl}\in\mathbb{C}^{N\times 1}$ is $\CN{\mathbf{0}}{\mathbf{R}_{kl}}$, $\mathbf{s} = [s_{1},s_{2},\ldots,s_{K}]^T \in \complexm{\numUEs}{1}$, $s_k$ is the transmit symbol (drawn from a constellation set, $\mathcal{S}$) of the $k$-th \gls{ue} with unit energy, and $\mathbf{n}_l$ is the noise at the receiver of \gls{ap} $l$ whose entries are independent and are distributed as circularly symmetric complex normal with mean $0$ and variance $\sigma^2$. {Hereafter, $\CN{\mathbf{0}}{\mathbf{C}}$ is used to denote a circularly symmetric complex normal distribution of mean $\mathbf{0}$ and covariance matrix $\mathbf{C}$ of appropriate dimensions.}

{\color{black}
We adopt analog communications to transmit the sufficient statistics of the \glspl{ue}' data from the \glspl{ap} to the \gls{cpu}, and propose an  \gls{ota} computation framework to combine them.
}


To motivate the \gls{ota} computation framework, we start with an expression for the \gls{ml} detector when the \gls{cpu} has obtained $\{\mathbf{y}_1, \ldots, \mathbf{y}_L\}$ and $\{\mathbf{H}_1,\ldots,\mathbf{H}_L\}$. Let us define the local sufficient statistics of the $l$-th \gls{ap} by 
{\color{black}
\begin{align}
\mathbf{\gram}_l&\triangleq\mathbf{H}_l^H\mathbf{H}_l=\begin{bmatrix}
     \norm{\mathbf{h}_{1l}}^2 & \mathbf{h}_{1l}^H\mathbf{h}_{2l} &\ldots & \mathbf{h}_{1l}^H\mathbf{h}_{Kl}\\
     \mathbf{h}_{2l}^H\mathbf{h}_{1l} & \norm{\mathbf{h}_{2l}}^2&\ldots&\mathbf{h}_{2l}^H\mathbf{h}_{Kl}\\
     \vdots & \vdots&\ddots&\vdots\\
     \mathbf{h}_{Kl}^H\mathbf{h}_{1l}&\mathbf{h}_{Kl}^H\mathbf{h}_{2l}&\ldots&\norm{\mathbf{h}_{Kl}}^2
 \end{bmatrix},  \label{eqn:GramMat1}\\
 \mathbf{\mf}_l &\triangleq \mathbf{H}^H_l\mathbf{y}_l.\label{eqn:MF_l}
\end{align}}
The Gramian matrix and the \gls{mf} output  are
\begin{align}
    \mathbf{\gram}&=\sum_{l=1}^{L}\mathbf{\gram}_l\in\mathbb{C}^{K\times K}\quad\text{and}\quad
\mathbf{\mf}=\sum_{l=1}^{L}\mathbf{\mf}_l\in\mathbb{C}^{K\times 1},\label{eq:SuffStats_mf}   
\end{align}
respectively. The \gls{ml} data detector is given by
\begin{align}
    \widehat{\mathbf{s}} 
    &=\argmin_{\mathbf{s}\in\mathcal{S}^K}\sum_{l=1}^L\norm{\mathbf{y}_l - \sqrt{p_{\rm ul}}\mathbf{H}_l\mathbf{s}}^2\nonumber\\
    &=\argmin_{\mathbf{s}\in\mathcal{S}^K}\, p_{\rm ul}\mathbf{s}^H \mathbf{\gram}\mathbf{s}-2\sqrt{p_{\rm ul}}\Re\left(\mathbf{s}^H\mathbf{\mf}\right).\label{eqn:SEROptMLDet}
\end{align}
From \eqref{eqn:SEROptMLDet}, we  see that it is sufficient for the \gls{cpu} to acquire only $\mathbf{\mf}$ and $\mathbf{\gram}$ to decode the data. Therefore, our goal is to develop a scheme that obtains $\mathbf{\gram}$ and $\mathbf{\mf}$ at the \gls{cpu} with less computational and communication overhead. Next we describe the \gls{ota} computation approach to achieve these  objectives.

\subsection{Two-Phase Transmission of the Sufficient Statistics}\label{sec:OTA_TwoPhasedTx}
Leveraging the observation that the sufficient statistics in \eqref{eq:SuffStats_mf} can be decomposed into a sum of local sufficient statistics available at the \glspl{ap}, we propose a two-phase approach to compute the sufficient statistics \gls{ota} using non-orthogonal resources. Accordingly, we divide the transmissions from the \glspl{ap} into two phases: the $l$-th \gls{ap} transmits $\mathbf{\gram}_{l}$ and $\mathbf{\mf}_{l}$ in the first and second phases, respectively.
We assume that $\numAntennasPerAP \geq \numAntennasCPU$ and that each \gls{ap} transmits $\numAntennasCPU$ symbols per channel use. Since $\mathbf{\gram}_{l}$ is a Hermitian matrix, it is enough to transmit either its upper or lower triangular part. Therefore, \gls{ap} $l$ transmits a vectorized form \colr{obtained by  row-wisely traversing the upper triangular part of} $\mathbf{\gram}_{l}$ in \eqref{eqn:GramMat1} denoted by $\mathbf{x}_{l}^{(1)}$ given in \eqref{eqn:Phase1Data} on the next page, which translates to $M_1 = \left\lceil \frac{K(K+1)}{2M} \right\rceil$ transmissions in the first phase. In the second phase, $M_2 = \left\lceil \frac{\tauu K}{M} \right\rceil$ transmissions are needed, where $\tauu$ is the number of channel uses by each \gls{ue} to transmit its \gls{ul} data. 
\begin{figure*}
    \begin{equation}
		\mathbf{x}_{l}^{(1)}=  
  {\begin{bmatrix}
				\norm{\mathbf{h}_{1l}}^2 & \mathbf{h}_{1l}^H\mathbf{h}_{2l} & \cdots & \mathbf{h}_{1l}^H\mathbf{h}_{Kl} & \norm{\mathbf{h}_{2l}}^2 & \mathbf{h}_{2l}^H\mathbf{h}_{3l}& \cdots & \mathbf{h}_{2l}^H\mathbf{h}_{Kl}& \cdots &
				\norm{\mathbf{h}_{Kl}}^2
			\end{bmatrix}^T}.  
 \label{eqn:Phase1Data}
\end{equation}
\hrule 
\vspace{-\baselineskip}
\end{figure*}
In the second phase, the $l$-th \gls{ap} transmits  
\begin{equation}
	\mathbf{x}_{l}^{(2)} = 
	\begin{bmatrix}
		\mathbf{\mf}_{l,1}^T  & \mathbf{\mf}_{l,2}^T  & \cdots & 
		\mathbf{\mf}_{l,\tauu}^T
	\end{bmatrix}^T,\label{eqn:Phase2Data}
\end{equation}
where the second index in the subscript of $\mathbf{\mf}_{l,t}$ denotes the $t$-th transmission interval of the \gls{mf} output $\mathbf{\mf}_{l}$. 

Let us denote the transmit signal matrix of \gls{ap} $l$ by
\begin{equation}
	\bar{\mathbf{X}}_{l}^{(i)} = 
	\begin{bmatrix} \bar{\mathbf{x}}_{l,1}^{(i)}&\ldots&\bar{\mathbf{x}}_{l,M_i}^{(i)}
	\end{bmatrix}
	\in\complexm{M}{M_i},\quad i\in\{1,2\},\label{eqn:Xbarmat}
\end{equation}
 where $\bar{\mathbf{x}}_{l,m}^{(i)}\in\complexm{M}{1}$ corresponds to the $((m-1)M+1)$-th to the $mM$-th entries of $\mathbf{x}_{l}^{(i)}$ given in \eqref{eqn:Phase1Data} (on top of \bluez{this} page) and \eqref{eqn:Phase2Data}. Note that $\bar{\mathbf{x}}_{l,M_i}^{(i)}$, $i\in\{1,2\}$ are zero padded if $M$ is not an integer multiple of $\frac{K(K+1)}{2}$ and $\tauu K$ for $i=1$ and $i=2$, respectively. We present the \gls{ota} framework to coherently combine the local sufficient statistics transmitted by the \glspl{ap} at the \gls{cpu}.

The received signal at the \gls{cpu} is given by
\begin{align}
	\mathbf{Z}^{(i)} = \sqrt{\eta_{\rm c}^{(i)}}\sum_{l=1}^L  \mathbf{G}_l^H\mathbf{W}_l         \bar{\mathbf{X}}_{l}^{(i)} + \mathbf{E}^{(i)}\in\complexm{M}{M_i},\label{eqn:RxSig1_CPU}
\end{align}
where $i\in\{1,2\}$ denotes the index of the two transmission phases of the \glspl{ap} to send their local sufficient statistics, $\mathbf{G}_l^H\in\complexm{M}{N}$ is the channel between the $l$-th \gls{ap} and the \gls{cpu}, $\mathbf{W}_l\in\complexm{N}{M}$ is the transmit precoder of the  $l$-th \gls{ap}, $\eta_{\rm c}^{(i)}$ is a common power assignment scaling factor of the \glspl{ap} to satisfy the average transmit power constraint
\begin{equation}\label{eq:APsPowConstraints}
    \eta_{\rm c}^{(i)} \expectL{\norm{\mathbf{W}_l\bar{\mathbf{X}}_{l}^{(i)}}^2_F} \leq P_{\rm max},
\end{equation}
where $\expectL{\cdot}$ represents the expectation of its argument with respect to all the associated random variables, $P_{\rm max}$ denotes the transmit power constraint for all the \glspl{ap}, and $\mathbf{E}^{(i)}$ is the additive noise at the \gls{cpu} with independent and identically entries distributed as $\CN{0}{\sigma^2}$~\cite{rouphael2014wireless}.


The \glspl{ap} obtain their fronthaul \gls{csi} to the \gls{cpu} using received pilot signals sent by the \gls{cpu}. \colr{Note that there is no need for pilot signals transmitted on orthogonal channels in order for the APs to obtain \gls{csi}. By exploiting reciprocity
of the AP-CPU channel, it is sufficient that the CPU broadcasts a pilot that is heard by all APs; based on this pilot, the APs estimate the CPU$\to$AP channels, which by virtue of reciprocity equal the APs$\to$CPU channels. } For ease of exposition, we assume that the fronthaul \gls{csi} available at the \glspl{ap} are perfect and have full column rank.
\footnote{In next-generation cell-free \gls{mimo} wireless systems, \glspl{ap} can be low-cost nodes strategically deployed by network planners. Consequently, the channels to the \gls{cpu} undergo rich scattering, supporting the assumption of full column rank.} 
We discuss the imperfect \gls{csi} scenario in Section~\ref{sec:impCSI}. Now to obtain the sum of the local sufficient statistics, each \gls{ap} first equalizes the effect of $\mathbf{G}_l$ using local \gls{zf} precoders ${\mathbf{W}_l = \mathbf{G}_l\left(\mathbf{G}_l^H \mathbf{G}_l\right)^{-1}\in\complexm{N}{M}}$, $l\in[L]$, where the $[\cdot]$ operator denotes the set of all integers from $1$ to its argument.

We recall that our goal is to compute $\sum_{l=1}^{L}\bar{\mathbf{X}}_{l}^{(i)}$, and although \gls{zf} precoding removes the effects of the channels between the \glspl{ap} and the \gls{cpu}, any unequal power scaling at the \glspl{ap} will result in residual scaling factors which are impossible to remove \gls{ota}. To circumvent this unequal power scaling problem, we propose an average transmit power assignment strategy at the \glspl{ap} to communicate with the \gls{cpu}.

The total energy expended by the $l$-th \gls{ap} during the $M_1$ and $M_2$ symbol intervals is
\begin{align}
	\Omega_l^{(i)} &= \eta_{\rm c}^{(i)}\,\sum_{t=1}^{M_i} \expectL{\norm{\mathbf{W}_l\bar{\mathbf{x}}_{l,t}^{(i)}}^2},\label{eqn:ForPowerConstraint}
\end{align}
where $i\in\{1,2\}$.\footnote{To obtain \eqref{eqn:ForPowerConstraint}, we assume that $\frac{K(K+1)}{2}$ and $\tauu K$ are multiples of $M$. However, we can handle the general case with a minor modification.} We use two separate scaling mechanisms for transmitting the Gramian and the \gls{mf} outputs from the \glspl{ap} to the \gls{cpu} to accommodate for the differences in their dynamic ranges. 
Note that the \gls{mf} output has $p_{\rm ul}$ embedded inside, which changes its dynamic range compared to that of the Gramian matrix. Therefore, it is imperative to employ this two-phase power assignment strategy.

The values of $\Omega_l^{(1)}$ and $\Omega_l^{(2)}$ can be computed using the corresponding mean and covariance matrices of the sufficient statistics which we will discuss subsequently. 
Therefore, the average transmit power of the $l$-th \gls{ap} to transmit the Gramian matrices and the \gls{mf} outputs of one coherence interval is
\begin{align}
	P_{l}^{(i)} = \frac{\Omega_l^{(i)}}{M_i},\qquad i\in\{1,2\}.\label{eqn:AvgTransPower}
\end{align}

We propose a low overhead feedback mechanism\footnote{\colr{The proposed feedback mechanism aligns with standardized uplink and downlink control signaling in \gls{lte} and 5G-NR~\cite{3GPP_TS_38_211}.}} to compute the common power scaling factors $\eta_{\rm c}^{(1)}$ and $\eta_{\rm c}^{(2)}$ at the \gls{cpu} (to assist in the coherent addition of the sufficient statistics at the \gls{cpu}) and broadcast it to the \glspl{ap} via a common control channel to satisfy their average transmit power constraints {\eqref{eq:APsPowConstraints}}. \colr{A small amount of control signaling between the CPU and APs is required (e.g., for power control purposes, when the large-scale fading changes) and  this signaling may advantageously take place on dedicated orthogonal channels. However, in practice, overall the amount of resources required for this will be small.}

For $i\in\{1,2\}$, each \gls{ap} sets $\eta_{\rm c}^{(i)}$ equal to $1$, computes the scalar value in \eqref{eqn:AvgTransPower} using the locally available statistics and conveys it to the \gls{cpu} using a dedicated control channel. 
This can be done after the pilot transmissions from the \gls{cpu} to the \glspl{ap}. Upon receiving $\{P_{1}^{(i)}, \ldots, P_{L}^{(i)}\}$, $i\in\{1,2\}$, the \gls{cpu} obtains the indices of the \glspl{ap} which violate the constraint. Let us include these indices in the sets $\mathcal{V}^{(i)} = \{i_1^{(i)}, \ldots, i_{L'^{(i)}}^{(i)}\}\subseteq\{1,\ldots,L\}$, where $L'^{(i)}$, $i\in\{1,2\}$, is the number of \glspl{ap} in it. Then the scaling factors are computed as
\begin{align}
    \eta_{\rm c}^{(i)} = \frac{P_{\rm max}}{\max_{j\in\mathcal{V}^{(i)}} P_{j}^{(i)}},\qquad i\in\{1,2\}.\label{eqn:computescalefactor1}
\end{align}

This ensures that every \gls{ap} satisfies its average transmit power constraint and is necessary to add the sufficient statistics coherently at the \gls{cpu}. Otherwise, the \gls{cpu} will receive an {unequally} weighted sum of sufficient statistics.

Substituting $\mathbf{W}_l = \mathbf{G}_l\left(\mathbf{G}_l^H \mathbf{G}_l\right)^{-1}$ into the received signal \eqref{eqn:RxSig1_CPU}, the \gls{cpu} receives the signal
\begin{equation}
	\mathbf{Z}^{(i)} \triangleq \begin{bmatrix}
		\mathbf{z}_1^{(i)}&\ldots&\mathbf{z}_{M_i}^{(i)}
	\end{bmatrix}= \sqrt{\eta_{\rm c}^{(i)}}\sum_{l=1}^{L}\bar{\mathbf{X}}_{l}^{(i)} + \mathbf{E}^{(i)},
\end{equation}
where $i\in\{1,2\}$, $\{\mathbf{z}_1^{(i)},\ldots,\mathbf{z}_{M_i}^{(i)}\}$ are the columns of $\mathbf{Z}^{(i)}$ and $\bar{\mathbf{X}}_{l}^{(i)}$ is given in \eqref{eqn:Xbarmat}. We define the \gls{snr} from the \glspl{ap} to the \gls{cpu} in two phases by
 \begin{align}
    \rho_{\rm c}^{(1)} &= \frac{\eta_c^{(1)} \expectL{\norm{\mathbf{\gram}}^2}}{K^2\sigma^2},\\
    \rho_{\rm c}^{(2)} &= \frac{\eta_c^{(2)} \expectL{\norm{\mathbf{\mf}}^2}}{K\sigma^2}.
\end{align}  

{\it Remark: }
We make an interesting remark that relates the transmit powers of the \glspl{ue} and $\eta_{\rm c}^{(2)}$. In a typical digital communication system, any multiuser detector performance always improves with the \gls{ul} transmit powers of \glspl{ue}. However, we observe a bottleneck in the \gls{ota} analog combining framework, which results in an error floor in \gls{ser} performance with respect to $p_{\rm ul}$. We first derive an analytical expression for $\eta_{\rm c}^{(2)}$.

{To see that, let us simplify the following term
\begin{equation}
    \begin{aligned}
        &\frac{1}{M_2}\sum_{t=1}^{M_2} \expectL{\norm{\mathbf{W}_l\bar{\mathbf{x}}_{l,t}^{(2)}}^2}\nonumber\\
    &=\frac{1}{M_2}\expectL{\norm{\begin{bmatrix}
	 \mathbf{W}_l & \mathbf{0}&\ldots &\mathbf{0}\\
     \mathbf{0}&\mathbf{W}_l &\ldots&\mathbf{0}\\
     \vdots&\vdots&\ddots&\vdots\\
     \mathbf{0}&\mathbf{0}&\ldots&\mathbf{W}_l
	\end{bmatrix}\begin{bmatrix}
	    \bar{\mathbf{x}}_{l,1}^{(2)}\\\vdots\\\bar{\mathbf{x}}_{l,M_2}^{(2)}
	\end{bmatrix}}^2}\\
    & =\frac{1}{M_2}\expectL{\norm{\widebar{\mathbf{W}}_l \mathbf{x}_{l}^{(2)}}^2} \\
    &=\frac{1}{M_2}\trace{\expectL{\widebar{\mathbf{W}}_l \mathbf{x}_{l}^{(2)}\mathbf{x}_{l}^{(2)H}\widebar{\mathbf{W}}_l^H}}\\
    &=\frac{1}{M_2}\left(p_{\rm ul}\trace{\expectL{\widebar{\mathbf{W}}_l\widebar{\mathbf{A}}_l^2\widebar{\mathbf{W}}_l^H}} + \sigma^2\trace{\expectL{\widebar{\mathbf{W}}_l\widebar{\mathbf{A}}_l\widebar{\mathbf{W}}_l^H}}\right)\\
    &= p_{\rm ul}a_l + b_l
    \end{aligned}
\end{equation}
where  $\widebar{\mathbf{A}}_l$ is a block-diagonal matrix with each block entry as $\mathbf{A}_l$, $\widebar{\mathbf{W}}_l=\mathbf{I}_{M_2}\otimes\mathbf{W}_l\in\mathbb{C}^{NM_2\times MM_2}$, $a_l \triangleq \frac{1}{M_2}\trace{\expectL{\widebar{\mathbf{W}}_l\widebar{\mathbf{A}}_l^2\widebar{\mathbf{W}}_l^H}} \geq 0$, and $b_l \triangleq \frac{\sigma^2}{M_2}\trace{\expectL{\widebar{\mathbf{W}}_l\widebar{\mathbf{A}}_l\widebar{\mathbf{W}}_l^H}}\geq 0$.

Now for a given $p_{\rm ul}$, we find
\begin{equation}
    r = \argmax_{l\in [L]} \, p_{\rm ul}a_l + b_l.
\end{equation}
Then, from \eqref{eqn:computescalefactor1}, we compute
\begin{align}
        \eta_{\rm c}^{(2)} &= \frac{P_{\rm max}}{p_{\rm ul}a_{r} + b_{r}},\label{eqn:eta_pul}
    \end{align}
}

From \eqref{eqn:eta_pul}, we see that: for a given $P_{\rm max}$, as $p_{\rm ul}$ increases, the common scaling factor $\eta_{\rm c}^{(2)}$ decreases. This results in a saturation of the received \gls{snr} of the sufficient statistics in the second phase as $p_{\rm ul}$ increases. Consequently, there will be an error floor in the \gls{ue} data detector's \gls{ser} performance. For example, for a given $P_{\rm max}$, the \gls{ser} of the \glspl{ue}' data does not decrease beyond a certain $p_{\rm ul}$. We discuss this error floor behavior further in detail in Section~\ref{sec:MSE_DataEst}.
In the next subsection, we provide the mean and covariance of the sufficient statistics, which are used to derive a Bayesian estimator of the sum of the sufficient statistics.

\subsection{Derivation of the Statistics of the Sufficient Statistics}\label{sec:SuffStatsDeriv}

As the channels between the \glspl{ap} and the \glspl{ue} are independent of each other, 
the mean $\boldsymbol{\mu}^{(1)}$ and the covariance matrix $\mathbf{C}^{(1)}$ of 
\begin{align}
    \mathbf{x}^{(1)}=\sum_{l=1}^L\mathbf{x}_{l}^{(1)}\label{eqn:x_1}
\end{align}
are given by
\begin{align}
    \boldsymbol{\mu}^{(1)} = \sum_{l=1}^L\boldsymbol{\mu}_{l}^{(1)}\qquad\text{and}\qquad
    \mathbf{C}^{(1)} =\sum_{l=1}^L\mathbf{C}_{l}^{(1)},\label{eqn:mu1C1stats}
\end{align}
respectively. Here $\boldsymbol{\mu}_{l}^{(1)}$ and $\mathbf{C}_{l}^{(1)}$ are the mean and covariance matrix of $\mathbf{x}_{l}^{(1)}$, $l\in [\numAPs]$, respectively. To compute $\boldsymbol{\mu}_{l}^{(1)}$ and $\mathbf{C}_{l}^{(1)}$ when the channels between the \glspl{ue} and the \glspl{ap} undergo correlated Rayleigh fading, we need the statistics of the sum of chi-squared and {the} product of complex Gaussian random variables. We compute them in closed form now. 

The nonzero entries of $\boldsymbol{\mu}_{l}^{(1)}$ and the diagonal entries of $\mathbf{C}_{l}^{(1)}$ are governed by the following indexing equations: For a  given $j\in[K]$ and $j' \in \{j,\ldots,K\}$, the non-zero $n$-th entry of $\boldsymbol{\mu}_{l}^{(1)}$ and $(n',n')$-th diagonal entry of $\mathbf{C}_{l}^{(1)}$ are at the indices ${n = (K-0.5j)(j-1) + j}$ and ${n' = (K-0.5j)(j-1)+j'}$, respectively, and are given by
\begin{align}
	\boldsymbol{\mu}_{l}^{(1)}[n] =\trace{{\mathbf{R}}_{jl}},\quad 
	\mathbf{C}_{l}^{(1)}[n',n'] = \trace{{\mathbf{R}}_{jl}{\mathbf{R}}_{j'l}},\ \label{eqn:C1l}
\end{align}
{where $\trace{\cdot}$ denotes the trace of matrix. 

For $i=2$, to compute the mean and covariance of each column in \eqref{eqn:Xbarmat}, we focus on the term $\sum_{l=1}^L\mathbf{\mf}_{l,t}^{(2)}$ i.e., the \gls{mf} output at time $t\in[\tau_u]$. For $t\neq t'$, $\mathbf{\mf}_{l,t}^{(2)}$ and $\mathbf{\mf}_{l,t'}^{(2)}$ are uncorrelated, and for $l\neq l'$, $\mathbf{\mf}_{l,t}^{(2)}$ and $\mathbf{\mf}_{l',t}^{(2)}$  are correlated. Therefore, to compute the mean and covariance of $\sum_{l=1}^L\mathbf{\mf}_{l,t}^{(2)}$, we need the mean $\boldsymbol{\mu}_{l,t}^{(2)}$, the covariance matrix $\mathbf{C}_{l,t}^{(2)}$ of $\mathbf{\mf}_{l,t}^{(2)}$ and also the cross-covariance matrix $\mathbf{C}_{ll',t}^{(2)}$ between $\mathbf{\mf}_{l,t}^{(2)}$ and $\mathbf{\mf}_{l',t}^{(2)}$. The mean is zero because $\boldsymbol{\mu}_{l,t}^{(2)}=\mathbf{0}$ for any $l$ and $t$. The covariance $\mathbf{C}^{(2)}$ of $\sum_{l=1}^L \mathbf{\mf}_{l,t}^{(2)}$ is given by
\begin{equation}
    \mathbf{C}^{(2)} = \sum_{l=1}^L\mathbf{C}_{l}^{(2)}   + \sum_{l=1}^L\sum_{\substack{l'=1,l'\neq l}}^L\mathbf{C}_{ll'}^{(2)}.\label{eqn:C2l1}
\end{equation}
For correlated Rayleigh fading, we {obtain}
\begin{align}
	\mathbf{C}_{l}^{(2)} &= p_{\rm ul}\expectL{\mathbf{\gram}_l^2}  + \sigma^2\expectL{\mathbf{\gram}_l},\label{eqn:C2l}\\
 \mathbf{C}_{ll'}^{(2)} &= p_{\rm ul}\expectL{\mathbf{\gram}_l}\expectL{\mathbf{\gram}_{l'}}.\label{eqn:C2lm}
\end{align}
where $\expect{\mathbf{\gram}_l} = {\rm diag}\left(\trace{{\mathbf{R}_{1l}}},\ldots,\trace{{\mathbf{R}}_{Kl}}\right)$, and $\expect{\mathbf{\gram}_l^2}$ is a diagonal matrix whose $k$-th diagonal entry is $\trace{{\mathbf{R}}_{kl}}^2 + \trace{{\mathbf{R}}_{kl}\sum_{k' = 1}^{K}{\mathbf{R}}_{k'l}}$. Finally, $\expect{\mathbf{W}_l^H\mathbf{W}_l}$ can be evaluated numerically. Based on the mean, covariance, and cross-covariance matrices of the sum of the sufficient statistics, we present the estimators below. 

\subsection{\Gls{lmmse} and \Acrshort{ls} Estimators}\label{sec:SuffStatsEstimators}
With a prior on the sufficient statistics, the \gls{cpu} considers a \gls{lmmse} estimator of $\sum_{l=1}^{L}\bar{\mathbf{X}}_{l}^{(i)}$, $i\in\{1,2\}$. For convenience, let us denote the mean and covariance of $\sum_{l=1}^{\numAPs}\bar{\mathbf{x}}_{l,m}^{(i)}$. $m\in[M_i],\ i\in\{1,2\}$ by $\bar{\boldsymbol{\mu}}_m^{(i)}$ and $\bar{\mathbf{C}}_m^{(i)}$, respectively. Note that $\bar{\boldsymbol{\mu}}_m^{(i)}$ and $\bar{\mathbf{C}}_m^{(i)}$ can be obtained by selecting the appropriate entries from the mean and covariance matrices derived in \eqref{eqn:mu1C1stats} and \eqref{eqn:C2l1}.
Then, the \gls{lmmse} estimate of $\sum_{l=1}^L \bar{\mathbf{x}}_{l,m}^{(i)}$, $i\in\{1,2\}$ (denoted by $\widehat{\bar{\mathbf{x}}}_m^{(i)}$), for $m\in[M_i],\ i\in\{1,2\}$ is
\begin{align}\label{eqn:LMMSEestimatSuffStats}
	\widehat{\bar{\mathbf{x}}}_m^{(i)} =\bar{\boldsymbol{\mu}}_m^{(i)} + \sqrt{\eta_{\rm c}^{(i)}}\bar{\mathbf{C}}_m^{(i)}&\left({\eta_{\rm c}^{(i)}}\bar{\mathbf{C}}_m^{(i)} + \sigma^2\mathbf{I}_{M}\right)^{-1}\nonumber\\
    &\times\left(\mathbf{z}_m^{(i)} - \sqrt{\eta_{\rm c}^{(i)}}\bar{\boldsymbol{\mu}}_m^{(i)}\right).
\end{align}

In the case when the \gls{cpu} does not have any prior information of the sufficient statistics, the \gls{cpu} implements the \gls{mvu} estimator
\begin{equation}\label{eqn:LSestimatSuffStats}
	\widehat{\bar{\mathbf{x}}}_m^{(i)} =\left({\eta_{\rm c}^{(i)}}\right)^{-\frac12}\mathbf{z}_m^{(i)},
\end{equation}
which is also efficient \cite{kay1993fundamentals}.  In this case, the \gls{mvu} estimator is also the \gls{ls} estimator. Finally, the \gls{cpu} computes the Gramian matrix and the \gls{mf} output estimates denoted by $\widehat{\mathbf{A}} = \widehat{\gramian{\mathbf{H}}}$ and $\widehat{\mathbf{\mf}} = \widehat{\mathbf{H}^H\mathbf{y}}$, respectively, by appropriately restructuring \eqref{eqn:LMMSEestimatSuffStats} or \eqref{eqn:LSestimatSuffStats}, and uses them to detect the \glspl{ue}' data.

\subsection{Data Detection}\label{sec:dataDetection}

In this section, we provide a few examples of the data estimators/detectors that use the obtained estimates of the sufficient statistics.  Note that our developed solution is equally applicable to any data detector.

\subsubsection{Linear Detectors}
The centralized \gls{lmmse} and \gls{ls} estimates of the data symbols are
\begin{equation}\label{eqn:approxLMMSE_Est}
    \begin{aligned}
    \widehat{\mathbf{s}}_{\text{\gls{lmmse}}} &=  \left(p_{\rm ul}\mathbf{\gram} + \sigma^2\mathbf{I}_{K} \right)^{-1}\mathbf{\mf}\\
        &\approx  \left(p_{\rm ul}\widehat{\mathbf{\gram}} + \sigma^2\mathbf{I}_{K} \right)^{-1}\widehat{\mathbf{\mf}},
    \end{aligned}    
\end{equation}
and 
\begin{equation}\label{eqn:approxLS_Est}
    \widehat{\mathbf{s}}_{\text{LS}} \approx p_{\rm ul}^{-1/2}\widehat{\mathbf{A}}^{-1}\widehat{\mathbf{\mf}}.
\end{equation}

We map the estimates from \eqref{eqn:approxLMMSE_Est} or \eqref{eqn:approxLS_Est} to the nearest constellation symbols to obtain the final detected symbols.

\subsubsection{\Gls{map} Detector}
The \gls{map} detector outputs
\begin{equation*}
   \begin{aligned}
        \widehat{\mathbf{s}}_{\text{MAP}} &=  \argmax_{\mathbf{s}\in \mathcal{S}} \norm{\mathbf{y} - \sqrt{p_{\rm ul}}\mathbf{H}\mathbf{s}}\\
        &= \argmax_{\mathbf{s}\in \mathcal{S}} \norm{\overline{\mathbf{y}} - \sqrt{p_{\rm ul}}\overline{\mathbf{H}}\mathbf{s}}\\
         &\approx  \argmax_{\mathbf{s}\in \mathcal{S}} \norm{\widehat{\overline{\mathbf{y}}} -\sqrt{p_{\rm ul}}\widehat{ \overline{\mathbf{H}}}\mathbf{s}},
   \end{aligned}
\end{equation*}
where ${\overline{\mathbf{H}}} = \mathbf{\gram}^{\frac{1}{2}}$, ${\overline{\mathbf{y}} }= \mathbf{\gram}^{-\frac{1}{2}}\mathbf{\mf}$, $\widehat{\overline{\mathbf{H}}} = \widehat{\mathbf{\gram}}^{\frac{1}{2}}$ and $\widehat{\overline{\mathbf{y}} }= \widehat{\mathbf{\gram}}^{-\frac{1}{2}}\widehat{\mathbf{\mf}}$ and $\mathcal{S}$ is the constellation set.

\subsubsection{Soft-output Detection}
We can also use soft detection methods to compute the bit \glspl{llr} as
\begin{equation}\label{eqn: llrSig}
    \mathcal{L}\left(b_i\rvert \widehat{\mathbf{\mf}}, \widehat{\mathbf{\gram}}\right) \approx \ln\left(\frac{\sum_{\mathbf{s}:b_i(\mathbf{s})=1}e^{-\frac{\norm{\widehat{\overline{\mathbf{y}}} - \sqrt{p_{\rm ul}}\widehat{\overline{\mathbf{H}}}\mathbf{s}}^2}{\sigma^2}}}{\sum_{\mathbf{s}:b_i(\mathbf{s})=0}e^{-\frac{\norm{\widehat{\overline{\mathbf{y}}} - \sqrt{p_{\rm ul}}\widehat{\overline{\mathbf{H}}}\mathbf{s}}^2}{\sigma^2}}}\right),
\end{equation}
where $\mathcal{L}(b_i) = \ln ({\rm Pr}(b_i=1)/{\rm Pr}(b_i=0))$, and the notation $\mathbf{s}:b_i(\mathbf{s})=\alpha$ denotes the set of all vectors $\mathbf{s}$ for which the $i$-th bit is $\alpha$. After computing the \glspl{llr}, we input them to a channel decoder to obtain the information bits.

\section{{Theoretical Performance Results}}
In this section, we present the theoretical results to analyze the performance of the \gls{ota} framework. We start with the \gls{mse} analysis of the estimator of the sufficient statistics followed by the achievable rate and the \gls{mse} of a \gls{ue}'s data.

\subsection{\gls{mse} of the Estimator of Sufficient Statistics}\label{MSE_SuffStats}
Our end goal is to detect the \glspl{ue}' data for which the accuracy of the estimates of the sufficient statistics is crucial. We derive analytical expressions of the \gls{mse} for the \gls{ls} and the \gls{lmmse} estimators of the Gramian and the \gls{mf} outputs. 

The \gls{mse} of the Gramian matrices that are transmitted in phase-$1$ as mentioned in Sec.~\ref{sec:OTA_TwoPhasedTx} is as follows:
\begin{equation}\label{eqn: mseSuffStats}
\expectL{\norm{\mathbf{\gram} - \widehat{\mathbf{\gram}}}_F^2} = 2\,\trace{\mathbf{C}_{e}^{(1)}} - \sum_{n\in \mathbb{J}}\mathbf{C}_{e}^{(1)}[n,n],
\end{equation}
where $\mathbf{C}_{e}^{(1)}$ is the error covariance of the estimator of $\mathbf{x}^{(1)}$ defined in \eqref{eqn:x_1}, $\mathbb{J} = \left\{(K-0.5k)(k-1) + k: \forall k\in [\numUEs]\right\}$ with cardinality $K$, and the expectation is with respect to all the associated random variables. 
For the \gls{ls} estimator, we compute the error covariance matrix as $\mathbf{C}_{e}^{(1)} = \frac{\sigma^2}{\eta_{\rm c}^{(1)}}\mathbf{I}$ which leads to
\begin{equation}\label{mse_p1_ls}
\expectL{\norm{\mathbf{\gram} - \widehat{\mathbf{\gram}}}_F^2}_{\rm LS} 
{=\left(\eta_{\rm c}^{(1)}\sigma^{-2}\right)^{-1}K^2},
\end{equation}
and for the \gls{lmmse} estimator
\begin{equation}\label{errcov_p1}
    \begin{aligned}
    \mathbf{C}_{e}^{(1)} &= \mathbf{C}^{(1)} - \eta_{\rm c}^{(1)}\mathbf{C}^{(1)}\left(\eta_{\rm c}^{(1)}\mathbf{C}^{(1)} + {\sigma^2}\mathbf{I}\right)^{-1}\mathbf{C}^{(1)},
    \end{aligned}
\end{equation}
where $\mathbf{C}^{(1)}$ is given in \eqref{eqn:mu1C1stats} \colr{ and is a diagonal matrix. This implies that the error covariance matrix in \eqref{errcov_p1} is also diagonal with its $(n',n')$-th entry given by 
\begin{equation}
  \mathbf{C}_{e}^{(1)}[n',n']  = \frac{\mathbf{C}^{(1)}[n',n']}{\frac{\eta_{\rm c}^{(1)}}{\sigma^2}\mathbf{C}^{(1)}[n',n'] + 1},
\end{equation}
where $\mathbf{C}^{(1)}[n',n'] = \sum_{l=1}^L\mathbf{C}_l^{(1)}[n',n']$. Therefore, the \gls{mse} of the Gramian with the \gls{lmmse} estimator is computed as
\begin{equation}\label{mse_p1_lmmse}
    \begin{aligned}
        &\expectL{\norm{\mathbf{\gram} - \widehat{\mathbf{\gram}}}_F^2}_{\rm LMMSE} \\
        &\,= 2\sum_{n^{'}=1}^{K(K+1)/2}\left(\mathbf{C}^{(1)}[n',n']^{-1} + \eta_{\rm c}^{(1)}\sigma^{-2}\right)^{-1} \\
        &\, \qquad 
        - \sum_{n\in \mathbb{J}}\left(\mathbf{C}^{(1)}[n,n]^{-1} + \eta_{\rm c}^{(1)}\sigma^{-2}\right)^{-1}.
    \end{aligned}
\end{equation}
}
\colr{From \eqref{mse_p1_ls} and \eqref{mse_p1_lmmse}, we see that the \gls{mse} of Gramian matrix decreases as $\eta_{\rm c}^{(1)}$ increases.}

For the \gls{ls} and \gls{lmmse} estimators, the \glspl{mse} of the \gls{mf} outputs are given by
\begin{align}
&\expectL{\norm{\mathbf{\mf} - \widehat{\mathbf{\mf}}}_F^2}_{\rm LS} 
\colr{=\left(\eta_{\rm c}^{(2)}\sigma^{-2}\right)^{-1}K},\label{eqn:mse_mfls}\\
&\expectL{\norm{\mathbf{\mf} - \widehat{\mathbf{\mf}}}_F^2}_{\rm LMMSE} =\sum_{i=1}^{K} \left(\mathbf{C}^{(2)}[i,i]^{-1} + \eta_{\rm c}^{(2)}\sigma^{-2}\right)^{-1},\label{eqn:mse_mflmmse}
\end{align}
respectively.

\colr{From \eqref{eqn:mse_mfls} and \eqref{eqn:mse_mflmmse}, we see that the \glspl{mse} of the \gls{mf} output with \gls{ls} and \gls{lmmse} estimators decreases as $\eta_{\rm c}^{(2)}$ increases. 
Alternatively, increasing $\eta_{\rm c}^{(2)}$ by lowering $p_{\rm ul}$, decreases the \gls{mse} in \eqref{eqn:mse_mflmmse}. Moreover, as $P_{\rm max}\rightarrow \infty$ (equivalently $\eta_{\rm c}^{(2)}\rightarrow \infty$), the \gls{mse} in all cases discussed above approaches zero.
}

\colr{For all the above estimators: (i) In principle, the \gls{mse} obtained by the \gls{lmmse} estimator equals at most that of the \gls{ls} estimator for both the Gramian and \gls{mf} outputs~\cite{kay1993fundamentals}. This is because
$\mathbf{C}^{(i)}[n',n']^{-1} + \eta_{\rm c}^{(i)}\sigma^{-2} > \eta_{\rm c}^{(i)}\sigma^{-2}$ for any $n'$, as $\mathbf{C}^{(i)}$ is a covariance matrix with positive diagonal entries. (ii) When $\eta_{\rm c}^{(i)}\sigma^{-2} \gg \mathbf{C}^{(i)}[n',n']^{-1}, \forall i$, the performance of the \gls{lmmse} estimator converges to that of the \gls{ls} estimator.}

\subsection{Derivation of Achievable Rate}
To evaluate the performance of the \glspl{ue}, we provide an expression for the achievable rate of the \glspl{ue}. We rewrite the output of a linear estimator as follows:
\begin{equation}\label{eqn:sigEstMethod1}
     \begin{aligned}
         \widehat{\mathbf{s}} &= \mathbf{V}_{\mathbf{s}}\widehat{\mathbf{H}^H\mathbf{y}}\\
     &= \mathbf{V}_{\mathbf{s}} \mathbf{V}_{\rm MF}{\mathbf{z}}^{(2)}\\
     &= \sqrt{\eta_{\rm c}^{(2)}p_{\rm ul}}\mathbf{V}_{\mathbf{s}} \mathbf{V}_{\rm MF}\mathbf{H}^H\mathbf{H}\mathbf{s} + \mathbf{V}_{\mathbf{s}} \mathbf{V}_{\rm MF}\left(\sqrt{\eta_{\rm c}^{(2)}}\mathbf{H}^H\mathbf{n} + \mathbf{e}\right)\\
     &= \sqrt{\eta_{\rm c}^{(2)}p_{\rm ul}}\mathbf{V}\mathbf{\gram}\mathbf{s} + \mathbf{V}\left(\sqrt{\eta_{\rm c}^{(2)}}\mathbf{H}^H\mathbf{n} + \mathbf{e}\right),
     \end{aligned}
\end{equation}
where $\mathbf{V}_{\mathbf{s}}$ is the linear receiver used to estimate the \glspl{ue}' data, $\mathbf{V}_{\rm MF}$ is the linear receiver to estimate the \gls{mf} output, $${\mathbf{z}}^{(2)} = \sqrt{\eta_{\rm c}^{(2)}}\mathbf{H}^H\mathbf{y} + \mathbf{e}^{(2)},\ \mathbf{e}^{(2)} \sim \CN{\mathbf{0}}{\sigma^2\mathbf{I}_M},$$ is the overall received signal at the \gls{cpu} during phase-2 and $\mathbf{V}\triangleq\mathbf{V}_{\mathbf{s}} \mathbf{V}_{\rm MF}$ is the effective data estimator. For instance,
\begin{equation}\label{eqn:V_LS1}
    \mathbf{V}_{\rm MF} = \left(\eta_{\rm c}^{(2)}\right)^{-\frac{1}{2}}\mathbf{I}_M,\  \mathbf{V}_{\mathbf{s}} = \left(p_{\rm ul}\right)^{-\frac{1}{2}}\widehat{\mathbf{A}}^{-1}
\end{equation}
is the effective \gls{ls} estimator.

We compute an achievable rate using the \gls{uatf} bound (see Section 2.3.4 of \cite{FundasMMIMO_2016}).\footnote{The actual achievable rate will be further higher, given that the \glspl{ap} and the \gls{cpu} have more knowledge of the CSI beyond just statistics. Nonetheless, formulating an expression for the achievable rate, contingent on such side information, is complex.} To compute this bound, we rewrite the received signal in \eqref{eqn:sigEstMethod1} as in \eqref{eqn:RecSigUtFb} (given on the top of the next page). Then, the achievable rate is given by
\begin{equation}\label{eqn:UatFb}
    {\rm R}_k^{\rm UatF} = \left(1 - \frac{\tau_p}{\tau_c}\right)\log_2 \left(1 + {\rm SINR}_k^{\rm UatF}\right)~{\rm bits/channel\,use},
\end{equation}
and ${\rm SINR}_k$ is given in \eqref{eqn:UtFbSINR}, where $\mathbf{a}_k$ and $\mathbf{v}_k$ are the $k$th columns of $\mathbf{\gram}$ and $\mathbf{V}^H$, respectively, and $\mathbf{v}_k$ is the receive combining vector {used} to estimate $s_k$. \colr{Note that, to achieve the rate in \eqref{eqn:UatFb}, the \gls{cpu} does not need to know the Gramian matrix $\mathbf{A}$, but only its first and second order statistics. The \gls{uatf}   bound is tight only when the 
channel hardens, or equivalently, the uncertainty in the effective channel gain  is small.
Generally, the bound is rather tight in massive-MIMO scenarios provided there are enough antennas; see for example \cite{bjornson2016massive} for a comparison between
empirical link performance using state-of-the-art modulation and coding to the
\gls{uatf} ergodic capacity bound. To assess the gap between the \gls{uatf} bound and ergodic capacity  in our system, we compared the \gls{uatf} SE  with  the SE obtained  based on the side-information ergodic bound for perfect \gls{csi}, where the Gramian and \gls{mf} output are known. This is shown in Fig. \ref{cdfSE}: the performance closely matches the perfect \gls{csi} case, showing that performance gap is minimal.}

\begin{figure*}[h!]
    \begin{align}
		&\widehat{s}_k = \sqrt{\eta_{\rm c}^{(2)}p_{\rm ul}}\expectL{\mathbf{v}_{k}^H\mathbf{a}_k}s_k + \sqrt{\eta_{\rm c}^{(2)}p_{\rm ul}}\left(\mathbf{v}_{k}^H\mathbf{a}_k-\expectL{\mathbf{v}_{k}^H\mathbf{a}_k}\right)s_k +\sqrt{\eta_{\rm c}^{(2)}p_{\rm ul}}\sum_{i\neq k}^{K}\mathbf{v}_k^H\mathbf{a}_i s_i + \sqrt{\eta_{\rm c}^{(2)}}\mathbf{v}_k^H\mathbf{H}^H\mathbf{n}+\mathbf{v}_k^H\mathbf{e}^{(2)}.\label{eqn:RecSigUtFb}\\
     &{\rm SINR}_k^{\rm UatF} = \frac{\rho_{\rm ul} \left\arrowvert{\expectL{\mathbf{v}_k^H\mathbf{a}_k}}\right\arrowvert^2}{\rho_{\rm ul}\sum_{i=1}^{K}\expectL{\left\arrowvert{\mathbf{v}_k^H\mathbf{a}_i}\right\arrowvert^2} - \rho_{\rm ul} \left\arrowvert{\expectL{\mathbf{v}_k^H\mathbf{a}_k}}\right\arrowvert^2 + \expectL{\norm{\mathbf{\gram}^{\frac{1}{2}}\mathbf{v}_k}^2} + \left(\eta_{\rm c}^{(2)}\right)^{-1}\expectL{\norm{\mathbf{v}_k}^2}}, \text{ where }\rho_{\rm ul} \triangleq \frac{p_{\rm ul}}{\sigma^2}.\label{eqn:UtFbSINR}
\end{align}
\hrule
\vspace{-\baselineskip}
\end{figure*}

\subsection{Mean-Square Error of the Users' Data Estimates
}\label{sec:MSE_DataEst}
{We emphasize that the sufficient statistics presented in \eqref{eq:SuffStats_mf} are valid regardless of the input signal's distribution~\cite{reeves2018mutual}. An important implication of this fact is that the proposed \gls{ota} framework is applicable even when the input signals are drawn from a continuous probability distribution, e.g., analog sensors' data. 
To evaluate the performance, one can compute the \gls{mse}. Moreover, the \gls{mse} serves as a crucial metric, providing a measure of the received \gls{snr}, for any input distribution}. The \gls{mse} of any generic linear estimator given in \eqref{eqn:sigEstMethod1} is given by
\begin{equation}\label{eqn:dataMSE}
    \begin{aligned}
		\expectL{\norm{\mathbf{s}-\widehat{\mathbf{s}}}^2}
		=&\expectL{\norm{\mathbf{I} - \mathbf{A}_1}^2_F} + \sigma^2\expectL{\norm{\mathbf{A}_2}_F^2 + \norm{\mathbf{V}}^2_F},
	\end{aligned}
\end{equation}
where $\mathbf{A}_1 = \sqrt{\eta_{\rm c}^{(2)}p_{\rm ul}}\mathbf{V}\mathbf{\gram}$, $\mathbf{A}_2 = \mathbf{V}\sqrt{\eta_{\rm c}^{(2)}}\mathbf{H}^H$.
In \eqref{eqn:dataMSE}, the first term on the right hand side is the error due to the residual of the channel after equalization, and the second term is the error due to overall noise.

\colr{
Recall that $\eta_{\rm c}^{(2)}$ (see \eqref{eqn:eta_pul}) is directly proportional to $P_{\rm max}$ and inversely proportional to $p_{\rm ul}$. We will analyze the impact of $P_{\rm max}$ and $p_{\rm ul}$ on the \gls{mse} of the \glspl{ue}' data. We discuss a case where the Gramian matrix $\mathbf{A}$ is perfectly estimated at the \gls{cpu} and use effective LS estimator. For this, $\mathbf{V}~=~\frac{1}{\sqrt{\eta_{\rm c}^{(2)}p_{\rm ul}}}\mathbf{A}^{-1}$ and accordingly we have
\begin{equation}\label{eqn:recsig_perZF}
    \widehat{\mathbf{s}}= \mathbf{s} + \underbrace{\frac{1}{p_{\rm ul}^{1/2}}\mathbf{A}^{-1}\mathbf{H}^H\mathbf{n} + \left(\frac{a_r}{P_{\max}} + \frac{b_r}{P_{\max} p_{\rm ul}}\right)^{1/2}\mathbf{A}^{-1}\mathbf{e}}_{\bar{\mathbf{n}}}.
\end{equation}

The \gls{mse} of the \glspl{ue} data becomes
\begin{equation}\label{eqn:dataMSE0}
    \begin{aligned}
         &\expectL{\norm{\mathbf{s}-\widehat{\mathbf{s}}}^2}
    =\\
    \, &\frac{\sigma^2}{p_{\rm ul}}\trace{\expectL{\mathbf{A}^{-1}}} + \left(\frac{a_r\sigma^2}{P_{\max}} + \frac{b_r\sigma^2}{P_{\max} p_{\rm ul}}\right)\trace{\expectL{\mathbf{A}^{-2}}},
    \end{aligned}
\end{equation}
and the following \gls{mse} asymptotic results hold:
\begin{equation}
    \begin{aligned}
        \lim_{p_{\rm ul}\rightarrow \infty}  \expectL{\norm{\mathbf{s}-\widehat{\mathbf{s}}}^2} &= \frac{\sigma^2}{P_{\rm max} }a_r\trace{\expectL{\mathbf{A}^{-2}}} > 0,\\
        \lim_{P_{\rm max}\rightarrow \infty}  \expectL{\norm{\mathbf{s}-\widehat{\mathbf{s}}}^2} &= \frac{\sigma^2}{p_{\rm ul}}\trace{\expectL{\mathbf{A}^{-1}}} > 0,\\
        \lim_{\substack{p_{\rm ul}\rightarrow\infty\\ P_{\rm max}\rightarrow \infty}}  \expectL{\norm{\mathbf{s}-\widehat{\mathbf{s}}}^2} &= 0.
    \end{aligned}
\end{equation}

We observe that if either $p_{\rm ul}$ or $P_{\rm max}$ are limited, the \gls{mse} does not decrease beyond a certain threshold. 
However, when both of them tend to $\infty$, the \gls{mse} becomes zero. Similar conclusions apply to the case when the Gramian estimate, i.e., $\mathbf{A}$, at the \gls{cpu} is imperfect, which also means that $P_{\rm max}$ is fixed (as with infinite $P_{\rm max}$, we will have a perfect estimation). In this case, an additional term appears, accounting for interference from other \glspl{ue}.

In \eqref{eqn:recsig_perZF} there are two  noise terms. Unless both $p_{\rm ul}$ and $P_{\rm max}$ are simultaneously increased, there will always be a nonzero interfering noise power that remains independent of one of the power parameters. 
To see that, consider the covariance of effective noise power, $\bar{\mathbf{n}}$ and the impact of power parameters on this:
\begin{equation}
   \expectL{\bar{\mathbf{n}}\bar{\mathbf{n}}^H}  = \frac{\sigma^2}{p_{\rm ul}}\expectL{\mathbf{A}^{-1}} + \sigma^2\left(\frac{a_r}{P_{\max}} + \frac{b_r}{P_{\max} p_{\rm ul}}\right)\expectL{\mathbf{A}^{-2}}.
\end{equation}
\begin{align}
    \lim_{p_{\rm ul} \rightarrow \infty}\expectL{\bar{\mathbf{n}}\bar{\mathbf{n}}^H} &= \frac{a_r\sigma^2}{P_{\max}}\expectL{\mathbf{A}^{-2}},\\
    \lim_{p_{\rm max} \rightarrow \infty}\expectL{\bar{\mathbf{n}}\bar{\mathbf{n}}^H} &= \frac{\sigma^2}{p_{\rm ul}}\expectL{\mathbf{A}^{-1}}.
\end{align}
This phenomenon leads to a \glspl{ue} performance floor. For instance, for a given $P_{\rm max}$, the \gls{ser} versus $p_{\rm ul}$ exhibits a floor beyond a certain $p_{\rm ul}$. We verify this phenomenon in the numerical results section, as shown in Fig.~\ref{plot_SER_UEs}. }


\subsection{{Alternative Method to Estimate \Glspl{ue}' Data } 
}
The data estimation framework presented in the {Section~\ref{sec:SystemModel}}
has two steps: {first, the Gramian and \gls{mf} output transmitted from the \glspl{ap} are estimated, and second, these estimates are used to detect \glspl{ue}'s data as in Section~\ref{sec:dataDetection}}. However, this two-step approach in the second phase can be combined into one single step and improve the performance. Specifically, note that the \gls{mf} output contains the information-bearing signals. Therefore, we can directly design the receive combining matrix $\mathbf{V}$ to estimate the {input symbols}, rather than the above mentioned two step approach. 
We can compute the \gls{ls} receive combining matrix as
\begin{equation}\label{eqn:V_LS2}
    \mathbf{V} =  \left(p_{\rm ul}\eta_{\rm c}^{(2)}\right)^{-\frac{1}{2}}\widehat{\mathbf{A}}^{-1}.
\end{equation}

To evaluate the performance of this approach, we calculate the achievable rate and the \gls{mse} using \eqref{eqn:UatFb} and \eqref{eqn:dataMSE}, respectively.

\subsection{Extension to Imperfect CSI}\label{sec:impCSI}
\subsubsection{Imperfect CSI between \Glspl{ue} and \Glspl{ap}}\label{sec:ImpCSI_APUE}
The \gls{ota} framework discussed in the previous sections assumes perfect \gls{csi} knowledge at the \glspl{ap}. In this section, we address the case when the \glspl{ap} have imperfect knowledge of the channels to the \glspl{ue}. To start with, each \gls{ap} obtains an \gls{lmmse} estimate of $\mathbf{H}_l, l \in [\numAPs]$, whose 
estimate and the estimation error are denoted by 
$\widehat{\mathbf{H}}_l$ and $\widetilde{\mathbf{H}}_l$, respectively. We rewrite the received signal at \gls{ap} $l$ given in \eqref{eqn:recSigAPs1} as
\begin{equation}
    \mathbf{y}_l = \sqrt{p_{\rm ul}}\widehat{\mathbf{H}}_l\mathbf{s} + \sqrt{p_{\rm ul}}\widetilde{\mathbf{H}}_l\mathbf{s}+ \mathbf{n}_l \in \complexm{N}{1}
\end{equation}

Note that for an \gls{mmse} estimator, the estimates and the estimation errors are uncorrelated, which we use to obtain the covariance matrix of the $l$-th column of $\widetilde{\mathbf{H}}_l$ i.e., $\widetilde{\mathbf{h}}_{kl}$ as
\begin{equation}
		\begin{aligned}
\widetilde{\mathbf{R}}_{kl} &= \expectL{\widetilde{\mathbf{h}}_{kl}\widetilde{\mathbf{h}}_{kl}^H}\\
  &= \mathbf{R}_{kl} - p_{\rm ul}\tau_p \mathbf{R}_{kl}\left(p_{\rm ul}\tau_p\sum_{i=1}^{K}\mathbf{R}_{il} + \sigma^2\mathbf{I}\right)^{-1}\mathbf{R}_{kl}.
		\end{aligned}
\end{equation}
Let us denote the covariance of $\sqrt{p_{\rm ul}}\widetilde{\mathbf{H}}_l\mathbf{s} + \mathbf{n}_l$ by $\boldsymbol{\Sigma}_l$. The \gls{ap} $l$ preprocesses the received signal as follows:
\begin{equation}
    \bar{\mathbf{y}}_l = \boldsymbol{\Sigma}_l^{-\frac{1}{2}}{\mathbf{y}}_l,
\end{equation}
where
\begin{equation}
\boldsymbol{\Sigma}_l = p_{\rm ul} \sum_{i=1}^{L}\widetilde{\mathbf{R}}_{il} + \sigma^2\mathbf{I}_{N}.
\end{equation}

The \gls{ap} $l$ then transmits $\mathbf{\gram}_l=\widehat{\mathbf{H}}_l^H\boldsymbol{\Sigma}_l^{-1}\widehat{\mathbf{H}}_l$ and $\mathbf{\mf}_l=\widehat{\mathbf{H}}_l^H\boldsymbol{\Sigma}_l^{-1}\mathbf{y}_l$, 
and the rest follows the same procedure as discussed in Section.~\ref{sec:OTA_TwoPhasedTx}. 
This information is approximately sufficient to decode the \glspl{ue}' data~\cite{shaik2021distributed}. The results presented for the perfect \gls{csi} can be adapted to imperfect \gls{csi} {by making appropriate changes.} 

\subsubsection{Imperfect CSI between \Glspl{ap} and \Gls{cpu}}\label{sec:ImpCSI_FH}
{\color{black} In  case there is imperfect knowledge of the channels between the \gls{cpu} and \glspl{ap}, residual errors may exist in the summed signal \gls{ota}. We briefly outline one potential way to handle this scenario. Suppose $\widehat{\mathbf{G}}_l$ and $\widetilde{\mathbf{G}}_l\triangleq \mathbf{G}_l-\widehat{\mathbf{G}}_l$ are the fronthaul \gls{csi} estimate and estimation error, respectively; when we apply local \gls{zf} precoding with imperfect fronthaul \gls{csi} at \glspl{ap}, the received signal (see \eqref{eqn:RxSig1_CPU}) at the \gls{cpu} becomes
\begin{align}
	\mathbf{Z}^{(i)} = &\sqrt{\eta_{\rm c}^{(i)}}\sum_{l=1}^L          \bar{\mathbf{X}}_{l}^{(i)} \nonumber\\&\,\,+ \sqrt{\eta_{\rm c}^{(i)}}\sum_{l=1}^L  \widetilde{\mathbf{G}}_l^H \widehat{\mathbf{G}}_l \left(\widehat{\mathbf{G}}_l^H \widehat{\mathbf{G}}_l\right)^{-1}         \bar{\mathbf{X}}_{l}^{(i)} + \mathbf{E}^{(i)},\label{eqn:RxSig1_CPUImpCSI}
\end{align}
where $i\in\{1,2\}$. With \gls{mmse} channel estimation, $\widehat{\mathbf{G}}_l$ and $\widetilde{\mathbf{G}}_l$, $l\in[L]$, are mutually uncorrelated. From the second term on the right-hand-side of \eqref{eqn:RxSig1_CPUImpCSI}, we see that residual errors due to channel imperfections contribute as an additive noise term, whose covariance matrix can be computed using the statistics of $\widehat{\mathbf{G}}_l$, $\widetilde{\mathbf{G}}_l$, and $\bar{\mathbf{X}}_{l}^{(i)}$ (derived in Section.~\ref{sec:SuffStatsDeriv}), $l\in [L]$. 
The rest follows as described in the previous sections. Note that there will be degradation in \gls{snr} due to imperfect fronthaul \gls{csi} at the \gls{cpu}, which leads to a loss in \gls{ser} performance. Another potential technique to mitigate degradation due to CSI imperfections is to increase the pilot length to estimate the channels between the \gls{cpu} and \glspl{ap}, though this comes with  additional overhead. We omit the details due to space limitations.}

\colr{\textit{Remark: } Note that the channel fading is a combination of two different effects: small-scale fading and large-scale fading.
The second effect, large-scale fading, is  influenced by path loss and shadowing. With OTA, the situation is similar to  in conventional cellular networks: some of the
links may be subject to very large path loss, and the weakest link will
eventually determine the SNR. To minimize the impact of this, APs should be strategically deployed in locations where path loss and shadowing effects are less severe, ensuring improved signal reliability and overall system performance.}

\section{Comparison of OTA with a Digital fronthaul}
The \gls{ota} computation discussed so far employs an analog summation of the signals transmitted from the \glspl{ap} to the \gls{cpu}. 
In this section, we address a crucial question: how does a digital transmission scheme compare to an \gls{ota} computation framework in terms of performance and complexity?

We consider an \gls{ods} where each \gls{ap} transmits on a non-overlapping set of resources using a conventional digital communication scheme. The \glspl{ap} quantize each of their complex source symbols to $2N_b$ bits ($N_b$ bits each to real and imaginary parts).\footnote{Note that this is a general representation of the signals transmitted by the \gls{ap}, which may include the \gls{mf} outputs, Gramian matrices, or any other signals.} We quantize the real and imaginary parts using floating point precision quantization and map them to a bit stream. We perform the channel coding and modulate its output to obtain the complex transmit symbols, which are sent via $N$ transmit antennas. 

The received signal at the \gls{cpu} is
\begin{equation}
	\mathbf{z} = \sum_{l=1}^{L}\chanMatCent{l}^H\boldsymbol{\theta}_l + \mathbf{e},\label{eqn:MAP_SysModel1}
\end{equation}
where $\boldsymbol{\theta}_l\in\mathbb{C}^{N\times 1}$ is the transmit signal of the $l$-th \gls{ap}, $\mathbf{e}\in \complexm{M}{1}$ is the \gls{cpu}'s additive noise {distributed as $\CN{\mathbf{0}}{\sigma^2\mathbf{I}}$.} 

A major bottleneck is that, when we employ an \gls{ods}, the number of channel uses needed to transmit all the symbols increases with the number of \glspl{ap}. Having $N_{s}$ complex source symbols translates to a total of $B = 2N_{s}N_b$ information bits  {per \gls{ap}}. If we denote the point-to-point \gls{mimo} channel capacity by $R_l$, then the total number of orthogonal resources (equivalently number of channel uses) needed is
\begin{equation}
    \Upsilon_l^{\rm \gls{ods}} \geq \ceil{\frac{B}{R_l}},
\end{equation}
where the achievable {ergodic} rate (assuming both the \gls{ap} $l$ and the \gls{cpu} have access to the corresponding fronthaul \gls{csi}) of \gls{ap} $l$ is given by

\begin{equation}\label{eq:ODSrate}
    \begin{aligned}
        R_l &= \mathbb{E}_{\mathbf{G}}\left\{\alpha_l\log_2\absL{\mathbf{I}_M + \mathbf{G}_l^H\mathbf{Q}_l\mathbf{G}_l}\right\},
    \end{aligned}
\end{equation}
where $\mathbf{Q}_l = \expectL{\boldsymbol{\theta}_l{\boldsymbol{\theta}_l}^H}$, and $\alpha_l \in [0,1]$ is the fraction of orthogonal resources assigned to the $l$-th \gls{ap} such that $\sum_{l=1}^{L}\alpha_l = 1$ and $\trace{\mathbf{Q}_l}\leq P_{\rm max}$. {This rate is achieved through} a waterfilling power allocation algorithm. 

The total number bits to share by each \gls{ap} is the same and equals $B$. So, $\{\alpha_l\}$ should be such that, all the \glspl{ap} have the same data rate, say $R$. To find $\{\alpha_l\}$, the following should hold
\begin{equation}
    \alpha_l \bar{R}_l = R, \forall l\in [L]
\end{equation}
where 
\begin{equation}
   \bar{R}_l =  \mathbb{E}_{\mathbf{G}}\left\{\log_2\absL{\mathbf{I}_M + \mathbf{G}_l^H\mathbf{Q}_l\mathbf{G}_l}\right\}.
\end{equation}

So considering the fact that $\sum_{l=1}^{L}\alpha_l = 1$, the common rate {is}
\begin{equation}
    R = \frac{1}{\sum_{l=1}^L\frac{1}{\bar{R}_l }}
\end{equation}
and the fraction of orthogonal resources assigned to AP $l$ is 
    $\alpha_l = \frac{R}{\bar{R}_l}.$ 
Then the overall number of orthogonal resources needed by \gls{ods} system is
\begin{equation}
    \begin{aligned}
        \Upsilon^{\rm \gls{ods}} &= \sum_{l=1}^{L}\Upsilon_l^{\rm \gls{ods}}
        \geq L\ceil{\frac{B}{R}}
    \end{aligned}
\end{equation}

For the \gls{ota} scheme, the number of channel uses needed to transmit $N_s$ symbols is
\begin{equation}\label{eqn:nbrChannUses}
    \Upsilon^{\rm OTA} = \ceil{\frac{N_s}{M}}
\end{equation}
by all the \glspl{ap}.

We see that with an increase in the number of \glspl{ap} and other system parameters fixed, the number of channel uses needed for \gls{ods} increases, which is not the case with \gls{ota}. 
Apart from the number of channel uses, the digital schemes also suffer from quantization errors which become prominent 
as the number of transmit nodes increases. The variance of the quantization error for the sum of the signals scales linearly with $L$ and $N_s$, which {can be} detrimental to the system performance. 


It is clear from the above discussion that \gls{ods} does not scale well with the number of \glspl{ap} as both the quantization error variance and the number of channel uses increase linearly with $L$. We numerically evaluate the performance of \gls{ods} in comparison with the \gls{ota} computation framework in the next section.

\section{Numerical Results}\label{sec:numericalResults}
In this section, we evaluate the performance of the proposed \gls{ota} framework numerically. We consider a simulation setup of $200~{\rm m}\times 200~{\rm m}$ where the \glspl{ue} are uniformly distributed, \glspl{ula} {are used} at the \glspl{ap}, and the \gls{cpu} is placed around the center at a height of $5~{\rm m}$. We adopt a 3GPP urban microcell channel propagation model with a carrier frequency of $2~{\rm GHz}$~\cite{LTE2010b}. We assume that the channels between the \glspl{ue} and the \glspl{ap}, and between the \glspl{ap} and the \gls{cpu} are spatially uncorrelated. The large-scale fading coefficient of the channel between the $k$-th \gls{ue} and the $l$-th \gls{ap} is calculated as: $$\beta_{kl}~[{\rm dB}]  = -30.5 -36.7\log_{10}({\rm d}_{kl}/1~{\rm m}),$$ where ${\rm d}_{kl}$ is the distance between the \gls{ue} $k$ and the \gls{ap} $l$. Note that $\beta_{kl}= {\rm tr}(\mathbf{R}_{kl})/N$. We set the other system parameters as $L=16$, $K=8$, $N=5$, and $M=4$. We use the same path loss model for the channels between the \glspl{ap} and \gls{cpu}.

\begin{figure}[!t] 
	\centering
	\begin{tikzpicture}
		\begin{axis}			
			[
			width=1\linewidth,
			height=0.85\linewidth,
			xmin = 0,
			xmax = 10,
			grid=both, 
			legend pos = north east,
			legend cell align={left},
			legend columns=1, 
			legend style={nodes={scale=0.75, transform shape}}, 
			xlabel = {$P_{\rm max}~[{\rm W}]$},
			ylabel = {\Gls{nmse} (dB)}
			]
			\addplot[thick, black, solid, mark = o,mark options={scale=2.0},each nth point=2] table [x index = 0, y index = 1, col sep=comma] {Figures/Data_MSE_Gramian_dB.dat};
			\addplot[thick, blue, solid,mark = square,mark options={scale=3.0},each nth point=2] table [x index = 0, y index = 2, col sep=comma] {Figures/Data_MSE_Gramian_dB.dat};

   \addplot[dashed, red, solid,mark = *,mark options={scale=2.0},each nth point=2] table [x index = 0, y index = 1, col sep=comma] {Figures/Data_MSE_MF_dB.dat};
   \addplot[dashed, red, solid,mark = o,mark options={scale=3.0},each nth point=2] table [x index = 0, y index = 2, col sep=comma] {Figures/Data_MSE_MF_dB.dat};
			%
			\addlegendentry{Sim. Gramian (\gls{lmmse}, \gls{ls})};  
                \addlegendentry{Theo. Gramian (\gls{lmmse}, \gls{ls})};

                \addlegendentry{Sim. \gls{mf} (\gls{lmmse}, \gls{ls})}; 
                \addlegendentry{Theo. \gls{mf} (\gls{lmmse}, \gls{ls})};
		\end{axis}
	\end{tikzpicture}
	\caption{\Gls{nmse} (dB) as a function of $P_{\rm max}$.}
	\label{plot_NMSE_SuffStats}
    \vspace{-0.9\baselineskip}
\end{figure}
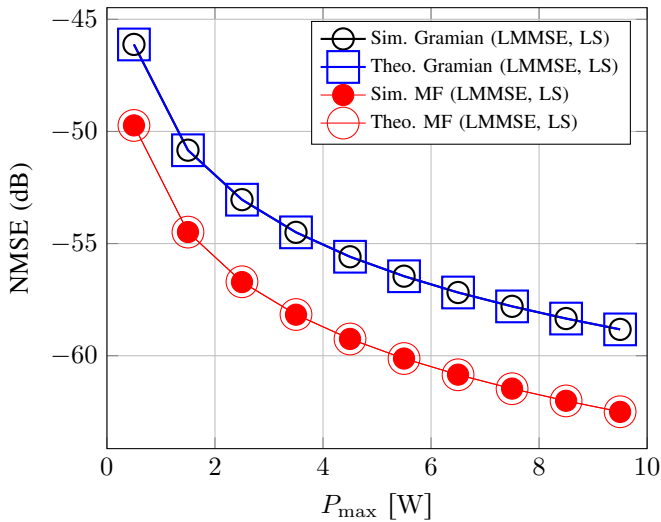
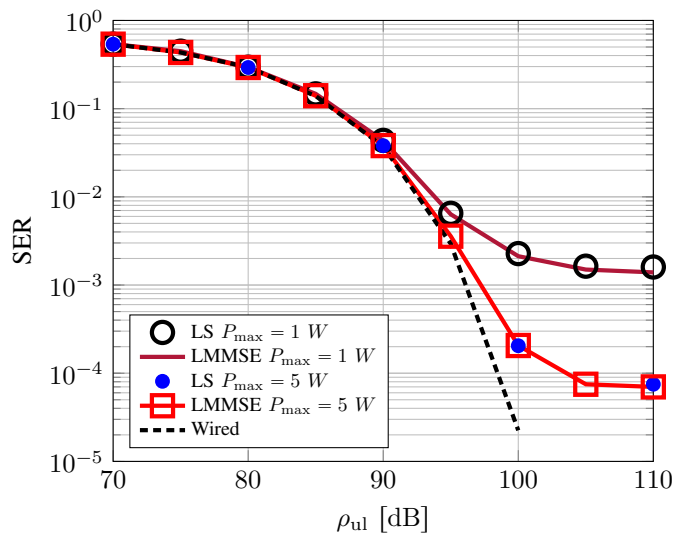
\begin{figure}[!t] 
	\centering
	\begin{tikzpicture}
		\begin{semilogyaxis}			
			[
			width=1\linewidth,
			height=0.85\linewidth,
			ymin = 1e-5,
			ymax = 1,			
			xmin = 70,
			xmax = 110,
			grid=both, 
			legend pos = south west,
			legend cell align={left},
			legend columns=1, 
			legend style={nodes={scale=0.75, transform shape}}, 
			xlabel = {$\rho_{\rm ul}~[{\rm dB}]$},
			ylabel = {SER}
			]
			\addplot[ultra thick, black, solid, mark = o,only marks,mark options={scale=2.0}] table [x index = 0, y index = 1, col sep=comma] {Figures/Data_SER_UEs.dat};
			\addplot[ultra thick,color={rgb:red,1;green,0.13;blue,0.32}, solid] table [x index = 0, y index = 2, col sep=comma] {Figures/Data_SER_UEs.dat};
   \addplot[thick, blue, solid, only marks, mark options={scale=1.2}, each nth point=2 ] table [x index = 0, y index = 3, col sep=comma] {Figures/Data_SER_UEs.dat};
   \addplot[ultra thick, red, solid, mark = square,mark options={scale=2}] table [x index = 0, y index = 4, col sep=comma] {Figures/Data_SER_UEs.dat};
   \addplot[ultra thick, black, densely dashed] table [x index = 0, y index = 6, col sep=comma] {Figures/Data_SER_UEs.dat};
			\addlegendentry{LS $P_{\max} = 1~W$};   
                \addlegendentry{LMMSE $P_{\max} = 1~W$};
                
                \addlegendentry{LS $P_{\max} = 5~W$};   
                \addlegendentry{LMMSE $P_{\max} = 5~W$};
                \addlegendentry{Wired}; 
		\end{semilogyaxis}
	\end{tikzpicture}
	\caption{SER as a function of $\rho_{\rm ul}$.}
 
	\label{plot_SER_UEs}
    \vspace{-0.9\baselineskip}
\end{figure}

In Fig.~\ref{plot_NMSE_SuffStats}, we plot the \gls{nmse} (defined as $\frac{\expect{\norm{\mathbf{x} - \widehat{\mathbf{x}}}^2}}{\expect{\norm{\mathbf{x}}^2}}$, where $\widehat{\mathbf{x}}$ is the estimate of a random vector $\mathbf{x}$) of the Gramian and the \gls{mf} outputs when the maximum transmit power of the \glspl{ap} is varied along the horizontal axis. 
We compare the \gls{nmse} between the theoretical and empirical values of the \gls{ls} and \gls{lmmse} estimators. This validates the accuracy of the closed form expressions derived in Sec.~\ref{MSE_SuffStats}. Further, we mention that the \gls{nmse} of the estimators is less than $-45$~dB 
even at a transmit power less than $0.5$~W, which demonstrates the efficacy of the proposed \gls{ota} method.

In Fig.~\ref{plot_SER_UEs}, we show the \gls{ser} performance of the \gls{lmmse} data detector as a function of $\rho_{\rm ul}$ when the \glspl{ue} transmit 4-\gls{qam} modulated symbols to the \glspl{ap}. The sufficient statistics are obtained using the \gls{ls} and \gls{lmmse} estimators given in \eqref{eqn:LMMSEestimatSuffStats} and \eqref{eqn:LSestimatSuffStats}, respectively, for $P_{\rm max}$ set to $\{1,5\}$~W. These estimates are used to compute the data of the \glspl{ue} using an \gls{lmmse} estimator given in \eqref{eqn:approxLMMSE_Est}, followed by a nearest-neighbor {data detector}. We observe that the \gls{ser} performance with the sufficient statistics estimated through the \gls{ls} and \gls{lmmse} estimators almost matches that of a cell-free massive \gls{mimo} system with a centralized data decoding. An interesting observation is that the \gls{ser} obtained by the \gls{ota} methods reaches an error floor beyond a particular value of $\rho_{\rm ul}$. Moreover, as $P_{\rm max}$ increases, the value of $\rho_{\rm ul}$ at which the error floor happens also increases. This is because, as $\rho_{\rm ul}$ increases, the scaling factor $\eta_{\rm c}^{(2)}$ to satisfy the average transmit power constraint during the transmission of the \gls{mf} outputs from the \glspl{ap} decreases. This leads to a saturation effect in the receive \gls{snr} at the \gls{cpu} resulting in an error floor. Further, as $P_{\rm max}$ increases, the saturation effect of the receive \gls{snr} at the \gls{cpu} occurs at a higher value of $\rho_{\rm ul}$. 
{One of the options to address this issue is to use a conventional digital communication scheme to transfer the local sufficient statistics from the \glspl{ap} with poor channel conditions. This approach improves the power scaling factors, leading to an increased receive \gls{snr} at the \gls{cpu}. Alternatively, these \glspl{ap} can be connected to the \gls{cpu} via a wired fronthaul, thereby eliminating the power scaling factor issues for them.} 

\begin{figure}[!t] 
	\centering
	\begin{tikzpicture}[spy using outlines={circle, magnification=6,connect spies}]
		\begin{axis}[
			width=1\linewidth,
			height=0.85\linewidth,
			ymin = 0,
			ymax = 1,			
			xmin =2,
			xmax = 14,
                grid=both,
			legend pos = north west,
			legend cell align={left},
			legend columns=1, 
			legend style={nodes={scale=0.75, transform shape},at={(0.4,0.3)},anchor=west}, 
			xlabel = {\Acrlong{se} (${\rm b/s/Hz}$)},
			ylabel = {CDF}
			]
			\addplot[ black, solid, each nth point=10] table [x index = 1, y index = 0, col sep=comma] {Figures/Data_cdfSEPow1.dat};
			\addplot[blue, solid, mark = square,mark options={scale=1},mark repeat=400,mark phase=1] table [x index = 2, y index = 0, col sep=comma] {Figures/Data_cdfSEPow1.dat};
			\addplot[ red, dashed, each nth point=10] table [x index = 3, y index = 0, col sep=comma]{Figures/Data_cdfSEPow1.dat};
   
            \addplot[ black, solid, each nth point=10, mark = o,mark repeat=50] table [x index = 1, y index = 0, col sep=comma] {Figures/Data_cdfSEPow2.dat};
			\addplot[blue, solid, , each nth point=10, mark = square*,mark options={scale=1},mark repeat=50,mark phase=1] table [x index = 2, y index = 0, col sep=comma] {Figures/Data_cdfSEPow2.dat};
			\addplot[ red, dashed, each nth point=10, mark = *,mark options={scale=1},mark repeat=50] table [x index = 3, y index = 0, col sep=comma]{Figures/Data_cdfSEPow2.dat};   
			\addlegendentry{OTA - \gls{uatf}, $P_\textrm{max} = 5~$W};
			\addlegendentry{Wired - \gls{uatf}, $P_\textrm{max} = 5~$W};			
			\addlegendentry{Wired - SI, $P_\textrm{max} = 5~$W};

            \addlegendentry{OTA - \gls{uatf}, $P_\textrm{max} = 0.1~$W};
			\addlegendentry{Wired - \gls{uatf}, $P_\textrm{max} = 0.1~$W};			
			\addlegendentry{Wired - SI, $P_\textrm{max} = 0.1~$W};
   \coordinate (spypoint) at (axis cs:4,0.13);
\coordinate (spyviewer) at (axis cs:4.5,0.75);
\spy[blue, size=2cm] on (spypoint) in node [fill=white] at (spyviewer);
		\end{axis}
	\end{tikzpicture}
	\caption{\Acrshort{cdf} of per user \acrlong{se}.}
	\label{cdfSE}
    \vspace{-0.9\baselineskip}
\end{figure}
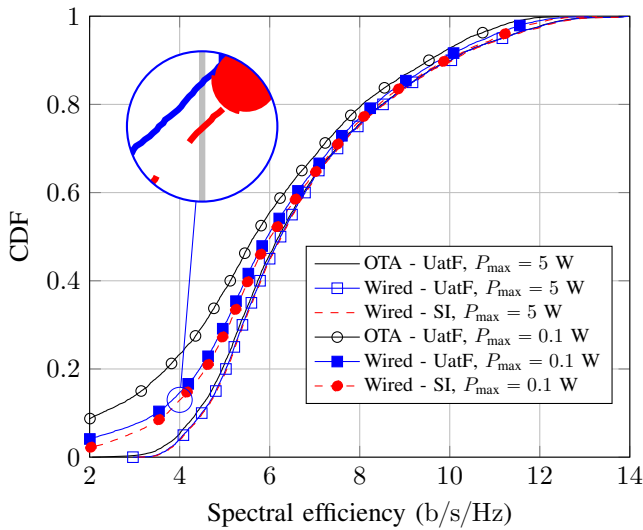

In Fig.~\ref{cdfSE}, we plot the \gls{cdf} of the per-user achievable rate obtained using the \gls{uatf} bound in \eqref{eqn:UatFb}. For this plot, we consider $\rho_{\rm ul} = 100~$dB, $P_{\rm max}=5~$W and  $P_{\rm max}=0.1~$W. We benchmark this result against the rate achieved with an ideal wired fronthaul between the \glspl{ap} and the \gls{cpu}, i.e., perfect channel equalization and a noise-free receiver at the \gls{cpu}. Mathematically, this is equivalent to $P_{\rm max}$ or $\eta_{c}^{(i)},\ i \in \{1,2\}$ tending to $\infty$ which translates to
\begin{equation}
    {\rm R}_k^{\rm Wired-UatF} = \lim_{\eta_{c}^{(2)}\rightarrow \infty} {\rm R}_k^{\rm UatF}.
\end{equation}

Another benchmark we compare with is the achievable rate obtained using the perfect \gls{csi} as side-information for the ideal wired fronthaul, which is given by~\cite{FundasMMIMO_2016}
\begin{equation}
    {\rm R}_k^{\rm Wired-SI} = \expectL{\log_2\left(1 + {\rm SINR}_k^{\rm SI}\right)},
\end{equation}
where
\begin{equation}
  {\rm SINR}_k^{\rm SI} = \frac{\rho_{\rm ul} \abs{\mathbf{v}_k^H\mathbf{a}_k}^2}{\rho_{\rm ul}\sum_{i=1}^{K}\abs{\mathbf{v}_k^H\mathbf{a}_i}^2  + \norm{\mathbf{H}\mathbf{v}_k}^2}
\end{equation}

From the \gls{cdf} plot of ${\rm R}_k^{\rm Wired-UatF}$ and ${\rm R}_k^{\rm Wired-SI}$, we observe that they almost overlap, which implies that there is sufficient channel hardening in the considered simulation setup to use the \gls{uatf} bound. Specifically, we note that the achievable rate of the \gls{ota} scheme is substantially close to that of the lossless transmission with $p_\textrm{ul}=5~$W. \colr{This implies that the rate expression in \eqref{eqn:UatFb} is indeed tight.} This further reaffirms that the \gls{ota} framework is not only scalable but also achieves almost the same performance as that of a wired fronthaul. To show the impact of $P_{\rm max}$ on different \acrlong{se} bounds clearly, we have also plotted {it} using $P_{\rm{max}}=0.1~$W. 

\begin{figure}[!t] 
	\centering
	\begin{tikzpicture}
		\begin{semilogyaxis}			
			[
			width=1\linewidth,
			height=0.85\linewidth,
			ymin = 1e-5,
			ymax = 1,			
			xmin = 60,
			xmax = 95,
			grid=both, 
			legend pos = south west,
			legend cell align={left},
			legend columns=1, 
			legend style={nodes={scale=0.75, transform shape}}, 
			xlabel = {$\rho_{\rm ul}~[{\rm dB}]$},
			ylabel = {BER}
			]
			\addplot[thick, black, solid, mark = o,mark options={scale=2.0},mark repeat=3,mark phase=1] table [x index = 0, y index = 1, col sep=comma] {Figures/Data_codedBER.dat};
			\addplot[ blue, mark = *,only marks,mark repeat=3,mark phase=1] table [x index = 0, y index = 2, col sep=comma] {Figures/Data_codedBER.dat};
   \addplot[thick, red, solid] table [x index = 0, y index = 1, col sep=comma] {Figures/Data_codedBER_refSNR.dat};
			\addlegendentry{OTA};   
                \addlegendentry{Wired};
                \addlegendentry{Ref. SNR}; 
		\end{semilogyaxis}
	\end{tikzpicture}
	\caption{BER as a function of $\rho_{\rm ul}$.}
	\label{plot_codedBER_UEs}
    \vspace{-0.9\baselineskip}
\end{figure}
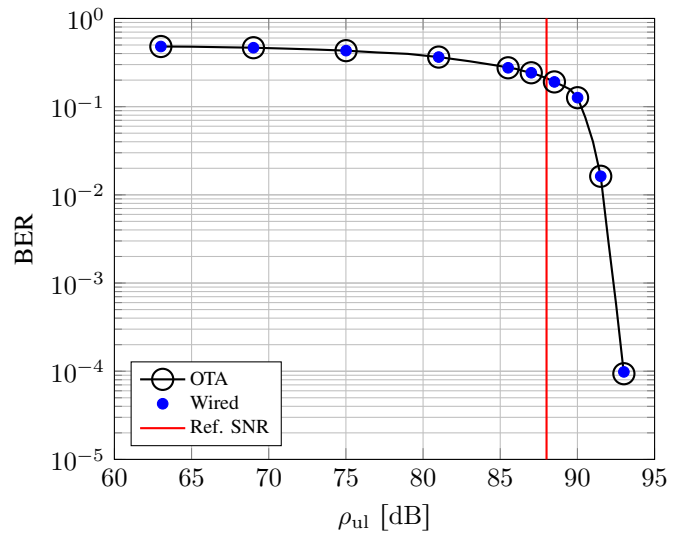

In Fig.~\ref{plot_codedBER_UEs}, we plot the coded \gls{ber} as a function of $\rho_{\rm ul}$ when $P_{\rm max}$ is set to $5$~W. We employ an LDPC error correcting code of rate ${\rm R_c} = 1/2$ and length $1944$ from the IEEE 802.11-2020 \gls{wlan} standard~\cite{standardIEEE802_11_2020}. We set $K$, $N$ and $M$ to $4$, $5$, and $3$, respectively. We used a max-log approximation of \eqref{eqn: llrSig} to compute the posterior bit \glspl{llr} and adopted the sphere decoding algorithm to obtain them efficiently. We have used the \gls{lmmse} estimator to obtain the sufficient statistics and benchmarked the coded \gls{ber} performance with a centralized wired fronthaul based system. We observe that channel coding not only reduces the performance gap between the wireless and wired fronthaul based systems, but also assists in mitigating the error floor issue seen in the Fig.~\ref{plot_SER_UEs}. The \gls{snr} gap to the achievable rate in \eqref{eqn:UatFb} is approximately $8~{\rm dB}$. Additionally, the \gls{se} bounds proposed in the paper are practically achievable, thereby serving as reliable indicators of overall link performance.

\subsection{Comparison of OTA with ODS}
In this subsection, we compare the end-to-end system performance between \gls{ods} and \gls{ota}. For \gls{ods}, we use a $N_b$ bit floating-point precision representation 
for the real and imaginary parts of transmit symbol with $1$ sign bit, $N_E$ exponent bits, and $N_F$ mantissa bits~\cite{IEEE754}. 

For the plots in the figures \ref{fig:plotNMSEvsNb}, \ref{fig:plotCUvsNb} and \ref{fig:plotSERvsRhoUL}, for a given $N_b$, we chose $N_E$ and $N_F$ to minimize the \gls{ser} of the \glspl{ue}. Additionally, for \gls{ods}, we set modulation and coding rate such that the net rate is below the achievable rate given in \eqref{eq:ODSrate}. For fairness, since \gls{ods} requires more channel uses than \gls{ota}, we repeat the \gls{ota} transmission on the extra channel uses. In other words, we increase the transmit power for \gls{ota} such that it mathces that of \gls{ods}. We set $K=8$, $N=5$, and $M=4$ for these plots.

In Fig.~\ref{fig:plotNMSEvsNb}, we show the \gls{nmse} of \gls{mf} output computed at the \gls{cpu} as a function of $N_b$ for \gls{ods} and \gls{ota} (corresponding to phase-$2$) with varying values of $L$. For \gls{ods}, as $N_b$ increases, the quantization error decreases which leads to a negative slope for the \gls{nmse} curve. We also observe that beyond $N_b=10$ bits
, the performance of \gls{ods} is similar for different values of $L$. \colr{In \gls{ods}, each \gls{ap} transmits its local sufficient statistics using orthogonal
	multiplexing and digital communications  with maximum permissible power. Further, for every AP, we employ capacity-achieving waterfilling-based power allocation. When the number of \glspl{ap} ($L$) increases and for a given \gls{adc} resolution ($N_b$), the quantization noise (to quantize the sufficient statistics) accumulated at the \gls{cpu} increases. Therefore, \gls{nmse} of the sufficient statistics does increase with $L$; however, the accumulated quantization noise is negligible when $N_b$ is greater than (say) $10$, and therefore the performance difference is marginal (for the particular parameters we have chosen).} We also see that \gls{ods} outperforms \gls{ota} after a certain number of quantization bits. This is because, in the case of \gls{ods}, each \gls{ap} operates with capacity-achieving point-to-point \gls{mimo} transmit and receive processing and water-filling based transmit power allocation. However, with \gls{ota}, each \gls{ap} may not transmit at full power to allow summation of transmitted signals coherently by all the \glspl{ap}. A question arising from the plot is whether \gls{ods} outperforms \gls{ota} after a certain number of quantization bits e.g., {$N_b=24$} bits for the presented simulation example. To answer this, in the next two plots we consider other performance metrics to gain insights into the network's end-to-end performance.

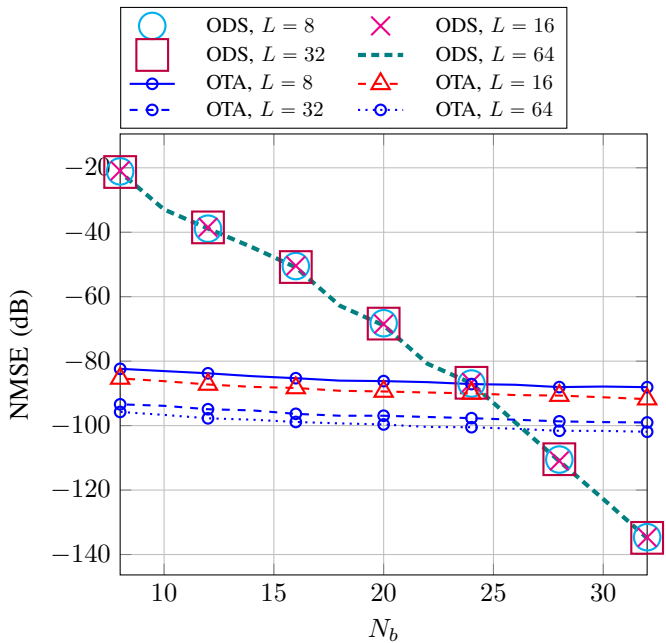
\begin{figure}[!t] 
	\centering
	\begin{tikzpicture}
		\begin{axis}			
			[
			width=0.98\linewidth,
			height=0.85\linewidth,
			xmin = 8,
			xmax = 32,
			grid=both, 
			legend columns=2, 
   legend style={nodes={scale=0.8, transform shape},at={(0,1.15)},anchor=west,column sep=10pt},
			legend cell align={left},
			xlabel = {$N_b$},
			ylabel = {\Gls{nmse} (dB)}
			]
			\addplot[thick,only marks,color=cyan, mark = o,mark options={solid},mark size = 5pt,mark repeat=2] table [x index = 0, y expr=10*log10(\thisrowno{2}), col sep=comma] {Figures/Data_nmseODSPhase2.dat};
            \addplot[only marks,thick, magenta, mark=x,mark options = solid,mark size = 5 pt, mark repeat=2] table [x index = 0, y expr=10*log10(\thisrowno{3}), col sep=comma] {Figures/Data_nmseODSPhase2.dat};
           \addplot[only marks,thick, purple, mark=square,mark options={solid},mark size = 6pt,mark repeat=2] table [x index = 0, y expr=10*log10(\thisrowno{4}), col sep=comma] {Figures/Data_nmseODSPhase2.dat};
            \addplot[ultra thick, teal, densely dashed] table [x index = 0, y expr=10*log10(\thisrowno{5}), col sep=comma] {Figures/Data_nmseODSPhase2.dat};
            \addplot[thick, blue, solid,mark=o,mark repeat=2] table [x index = 0, y expr=10*log10(\thisrowno{2}), col sep=comma] {Figures/Data_nmseOTAPhase2.dat};
            \addplot[thick, red, dashed,mark=triangle,mark options={solid},mark size = 4pt,mark repeat=2] table [x index = 0, y expr=10*log10(\thisrowno{3}), col sep=comma] {Figures/Data_nmseOTAPhase2.dat};
            \addplot[thick, blue, dashed,mark=o,mark options={solid},mark repeat=2] table [x index = 0, y expr=10*log10(\thisrowno{4}), col sep=comma] {Figures/Data_nmseOTAPhase2.dat};
            \addplot[thick, blue, dotted,mark=o,mark options={solid},mark repeat=2] table [x index = 0, y expr=10*log10(\thisrowno{5}), col sep=comma] {Figures/Data_nmseOTAPhase2.dat};
			\addlegendentry{ODS, $L=8$};
			\addlegendentry{ODS, $L=16$};			
			\addlegendentry{ODS, $L=32$};
            \addlegendentry{ODS, $L=64$};
   
			\addlegendentry{OTA, $L=8$};		
			\addlegendentry{OTA, $L=16$};
			\addlegendentry{OTA, $L=32$};
           \addlegendentry{OTA, $L=64$};
		\end{axis}
	\end{tikzpicture}
	\caption{\gls{nmse} (dB) of \gls{mf} output as a function of $N_b$ with $\rho_\textrm{ul}=110~\textrm{dB}$ and $P_\textrm{max} = 5~W$.}
 \label{fig:plotNMSEvsNb}
 \vspace{-0.9\baselineskip}
\end{figure}
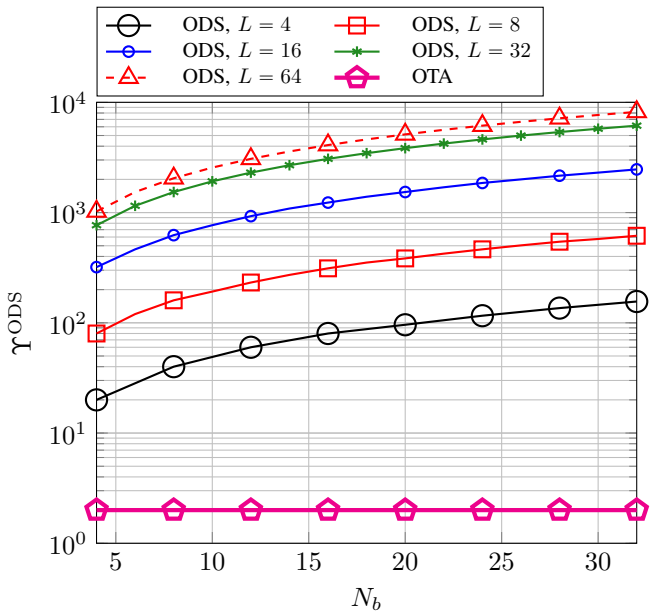
\begin{figure}[!t] 
	\centering
	\begin{tikzpicture}
		\begin{semilogyaxis}			
			[
			width=1\linewidth,
			height=0.85\linewidth,
			ymin = 1,
			ymax = 1e4,			
			xmin = 4,
			xmax = 32,
			grid=both, 
			legend pos = north east,
			legend cell align={left},
			legend columns=2, 
			legend style={nodes={scale=0.8, transform shape},at={(0,1.115)},anchor=west,column sep=10pt},
			xlabel = {$N_b$},
			ylabel = {$\Upsilon^{\rm \gls{ods}}$}
			]
			\addplot[thick, black, solid, mark = o,mark options={scale=2.0},each nth point=2] table [x index = 0, y index = 1, col sep=comma] {Figures/Data_ChanUses_ODS_OTA_Phase2.dat};
			\addplot[thick, red, solid, mark=square,mark size = 3pt,mark repeat=2] table [x index = 0, y index = 2, col sep=comma] {Figures/Data_ChanUses_ODS_OTA_Phase2.dat};
            \addplot[thick, blue, solid,mark=o,mark repeat=2] table [x index = 0, y index = 3, col sep=comma] {Figures/Data_ChanUses_ODS_OTA_Phase2.dat};
            \addplot[thick, forestgreen, solid,mark=asterisk,mark size=2pt, mark options={solid}] table [x index = 0, y index = 4, col sep=comma] {Figures/Data_ChanUses_ODS_OTA_Phase2.dat};
            \addplot[thick, red, dashed,mark=triangle,mark options={solid},mark size = 4pt,mark repeat=2] table [x index = 0, y index = 5, col sep=comma] {Figures/Data_ChanUses_ODS_OTA_Phase2.dat};
            \addplot[ultra thick, magenta, solid,mark=pentagon,mark size=4, mark repeat = 2] table [x index = 0, y index = 6, col sep=comma] {Figures/Data_ChanUses_ODS_OTA_Phase2.dat};

		\addlegendentry{ODS, $L=4$};  
            \addlegendentry{ODS, $L=8$}; 
            \addlegendentry{ODS, $L=16$};  
            \addlegendentry{ODS, $L=32$}; 
            \addlegendentry{ODS, $L=64$}; 

            \addlegendentry{OTA}; 
		\end{semilogyaxis}
	\end{tikzpicture}
	\caption{$\Upsilon^{\rm \gls{ods}}$ (for phase-$2$) as a function of  $N_b$ with $\rho_\textrm{ul}=110~\textrm{dB}$ and $P_\textrm{max} = 5~W$.}
    \label{fig:plotCUvsNb}
    \vspace{-0.9\baselineskip}
\end{figure}

In Fig.~\ref{fig:plotCUvsNb}, we plot the number of channel uses $\Upsilon^{\rm \gls{ods}}$ (to transmit the \gls{mf} output) as a function of $N_b$. We observe that $\Upsilon^{\rm \gls{ods}}$ scales linearly with $N_b$ for \gls{ods}, whereas it is constant for \gls{ota}. Moreover, \gls{ota} needs much {fewer} number of channel uses than \gls{ods}. This highlights one of the benefits of the \gls{ota} framework compared to \gls{ods} as the number of \glspl{ap} become large. Moreover, this effect may also lead to a memory buffer overload at the \glspl{ap} for \gls{ods}.
\begin{figure}[!t] 
	\centering
	\begin{tikzpicture}
		\begin{semilogyaxis}			
			[
			width=1\linewidth,
			height=0.8\linewidth,
			ymin = 1,
			ymax = 1e4,			
			xmin = 4,
			xmax = 64,
			xmode=log, 
			log basis x=2, 
			grid=both, 
			legend pos = north east,
			legend cell align={left},
			legend columns=2, 
			legend style={nodes={scale=0.8, transform shape},at={(0,1.115)},anchor=west,column sep=10pt},
			xlabel = {$L$},
			ylabel = {$\Upsilon^{\rm \gls{ods}}$},
			xtick={4,8,16,32,64}, 
			xticklabels={$4$, $8$, $16$, $32$, $64$} 
			]
			\addplot[thick, black, solid, mark = o,mark options={scale=2.0}] table [x index = 0, y index = 1, col sep=comma] {Figures/Data_ChanUses_Vs_L_ODS_OTA_Phase2.dat};
            
			\addplot[thick, red, solid, mark=square,mark size = 3pt] table [x index = 0, y index = 3, col sep=comma] {Figures/Data_ChanUses_Vs_L_ODS_OTA_Phase2.dat};
            
            \addplot[thick, blue, solid,mark=o] table [x index = 0, y index = 6, col sep=comma] {Figures/Data_ChanUses_Vs_L_ODS_OTA_Phase2.dat};
            
            \addplot[thick, forestgreen, solid,mark=asterisk,mark size=2pt, mark options={solid}] table [x index = 0, y index = 15, col sep=comma] {Figures/Data_ChanUses_Vs_L_ODS_OTA_Phase2.dat};
            
            \addplot[thick, red, dashed,mark=triangle,mark options={solid},mark size = 4pt] table [x index = 0, y index = 16, col sep=comma] {Figures/Data_ChanUses_Vs_L_ODS_OTA_Phase2.dat};

		  \addlegendentry{ODS, $N_b=4$};  
            \addlegendentry{ODS, $N_b=8$}; 
            \addlegendentry{ODS, $N_b=16$};  
            \addlegendentry{ODS, $N_b=32$}; 

            \addlegendentry{OTA}; 
		\end{semilogyaxis}
	\end{tikzpicture}
	\caption{$\Upsilon^{\rm \gls{ods}}$ (for phase-$2$) as a function of  $L$ with $\rho_\textrm{ul}=110~\textrm{dB}$ and $P_\textrm{max} = 5~W$.}
    \label{fig:plotCUvsL22}
    \vspace{-0.9\baselineskip}
\end{figure}
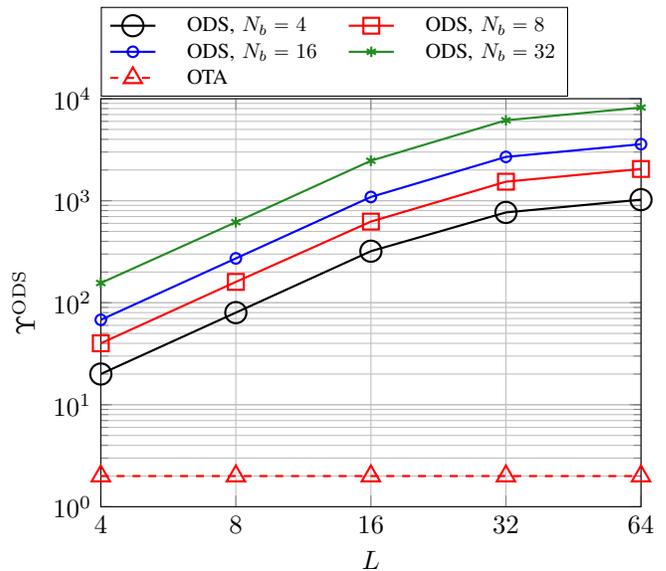

\begin{figure}[!t] 
	\centering
	\begin{tikzpicture}
		\begin{semilogyaxis}			
			[
			width=1\linewidth,
			height=0.8\linewidth,
			ymin = 1e-3,
			ymax = 1,			
			xmin = 70,
			xmax = 110,
			grid=both, 
			legend pos = south west,
			legend cell align={left},
			legend columns=1, 
			legend style={nodes={scale=0.75, transform shape}}, 
			xlabel = {$\rho_{\rm ul}~[{\rm dB}]$},
			ylabel = {SER}
			]
			\addplot[thick, forestgreen, solid,mark=pentagon,mark size=2pt, mark options={solid}] table [x index = 0, y index = 1, col sep=comma] {Figures/Data_ODS_SER_Nb.dat};
			\addplot[ultra thick,color={rgb:red,1;green,0.13;blue,0.32}, solid] table [x index = 0, y index = 1, col sep=comma] {Figures/Data_OTAnoExtraSNR_SER_Nb.dat};
            \addplot[thick, red, solid, mark=square,mark size = 3pt,mark repeat=2] table [x index = 0, y index = 2, col sep=comma] {Figures/Data_ODS_SER_Nb.dat};
            \addplot[thick, blue, solid,mark=o,mark repeat=2] table [x index = 0, y index = 1, col sep=comma] {Figures/Data_OTA_SER_Nb.dat};
            \addplot[thick, forestgreen, solid,mark=asterisk,mark size=2pt, mark options={solid}] table [x index = 0, y index = 2, col sep=comma] {Figures/Data_OTA_SER_Nb.dat};
            \addplot[thick, black, solid] table [x index = 0, y index = 1, col sep=comma] {Figures/Data_Genie_SER_Nb.dat};
			\addlegendentry{ODS $N_b = 8$};   
            \addlegendentry{OTA No Extra SNR};   
                
            \addlegendentry{ODS $N_b = 16$};
            \addlegendentry{OTA $N_b = 8$}; 
            \addlegendentry{OTA $N_b = 16$}; 
            \addlegendentry{Genie}; 
		\end{semilogyaxis}
	\end{tikzpicture}
	\caption{\acrshort{ser} as a function of $\rho_{\rm ul}$ with $P_\textrm{max} = 5~W$ and $L=16$.}
    \label{fig:plotSERvsRhoUL}
    \vspace{-0.9\baselineskip}
\end{figure}
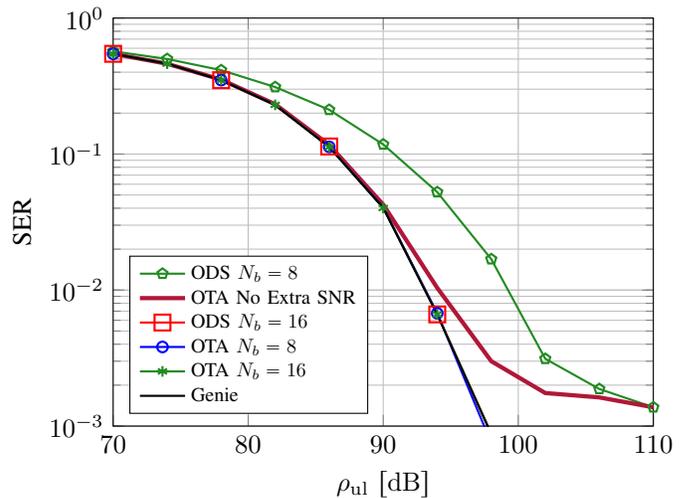

In Fig.~\ref{fig:plotSERvsRhoUL}, we plot the \gls{ser} of the \glspl{ue} comparing \gls{ods} and \gls{ota} schemes. We observe that \gls{ods} requires at least $N_b = 16$ bits per real symbol to match the performance of \gls{ota}. Additionally, the performance of \gls{ota} without the extra \gls{snr} i.e., ratio of $\Upsilon^{\rm \gls{ods}}$ and $\Upsilon^{\rm \gls{ota}}$, due to additional channel uses is still comparable to the \gls{ods} method up to an \gls{snr} of around $98~$dB. We also observe that \gls{ods} with $N_b = 8$ has the poorest performance among all the methods. These observations indicate that \gls{ota} is a suitable choice for end-to-end \glspl{ue} performance, requiring the least number of channel uses, which would otherwise necessitate a significantly larger number of channel uses and high resolution \glspl{adc}. 


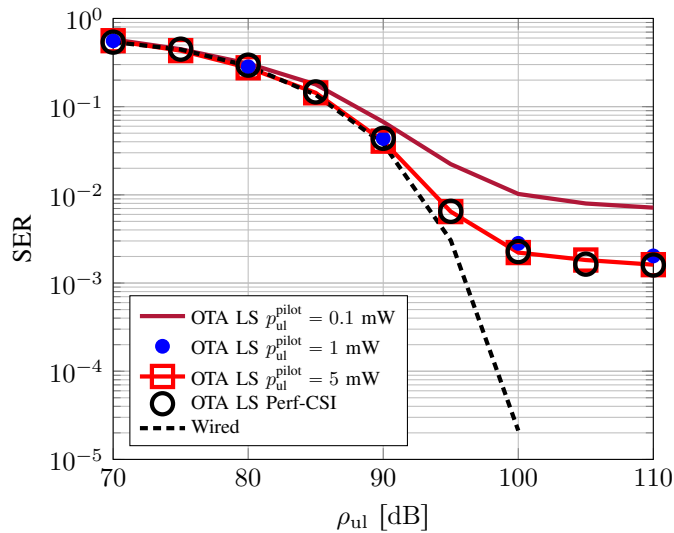
\begin{figure}[!t] 
	\centering
	\begin{tikzpicture}
		\begin{semilogyaxis}			
			[
			width=1\linewidth,
			height=0.85\linewidth,
			ymin = 1e-5,
			ymax = 1,			
			xmin = 70,
			xmax = 110,
			grid=both, 
			legend pos = south west,
			legend cell align={left},
			legend columns=1, 
			legend style={nodes={scale=0.75, transform shape}}, 
			xlabel = {$\rho_{\rm ul}~[{\rm dB}]$},
			ylabel = {\Acrshort{ser}}
			]
			\addplot[ultra thick,color={rgb:red,1;green,0.13;blue,0.32}, solid] table [x index = 0, y index = 1, col sep=comma] {Figures/Data_SER_UEs_ImpCSI.dat};
            \addplot[thick, blue, solid, only marks, mark options={scale=1.2}, each nth point=2 ] table [x index = 0, y index = 2, col sep=comma] {Figures/Data_SER_UEs_ImpCSI.dat};
            \addplot[ultra thick, red, solid, mark = square,mark options={scale=2}] table [x index = 0, y index = 3, col sep=comma] {Figures/Data_SER_UEs_ImpCSI.dat};   
			\addplot[ultra thick, black, solid, mark = o,only marks,mark options={scale=2.0}] table [x index = 0, y index = 1, col sep=comma] {Figures/Data_SER_UEs.dat};
            \addplot[ultra thick, black, densely dashed] table [x index = 0, y index = 4, col sep=comma] {Figures/Data_SER_UEs_ImpCSI.dat};
            \addlegendentry{OTA LS $p_\textrm{ul}^\textrm{pilot} = 0.1~$mW};   
            \addlegendentry{OTA LS $p_\textrm{ul}^\textrm{pilot} = 1~$mW};   
            \addlegendentry{OTA LS $p_\textrm{ul}^\textrm{pilot} = 5~$mW};  
            \addlegendentry{OTA LS Perf-CSI};    
            \addlegendentry{Wired}; 
		\end{semilogyaxis}
	\end{tikzpicture}
	\caption{SER as a function of $\rho_{\rm ul}$ with imperfect \gls{csi} at the \glspl{ap}.}
 
	\label{plot_SER_UEs_ImpCSI}
    \vspace{-0.9\baselineskip}
\end{figure}

In Fig.~\ref{plot_SER_UEs_ImpCSI}, we plot \gls{ser} versus $\rho_{\rm ul}$ for imperfect \gls{csi} with different pilot powers, denoted by $p_\textrm{ul}^\textrm{pilot}$ in the figure. We observe through simulations that the performance with imperfect \gls{csi} is comparable to that of perfect \gls{csi} for $p_\textrm{ul}^\textrm{pilot} = 1~$mW. This implies that the approximate sufficient statistics for imperfect \gls{csi} described in Section~\ref{sec:impCSI} serve as a reasonable proxy for the exact sufficient statistics.  

\colr{Furthermore, while imperfect \gls{csi} affects, with proper pilot power allocation, the resulting performance degradation remains minimal with the chosen system parameters. This suggests that the proposed approach is robust to small estimation errors.}

\colr{Note that we used \gls{zf} local precoding for the sufficient statistics. It is plausible that a small performance improvement could be achieved using MMSE (or similar) instead of ZF. However, our simulation results do show that UE performance is close to that in the wired case, demonstrating that ZF is sufficient at least in the scenarios 
considered.}

\section{Conclusions}\label{sec:conclu}

{We developed a novel and scalable technique for wireless fronthauling in \gls{ul} cell-free massive \gls{mimo} systems, exploiting  \gls{ota} computation  to obtain the global sufficient statistics needed to decode the \glspl{ue}' data at the \gls{cpu}. Specifically, we provided a two-phase mechanism and a power assignment strategy to coherently combine the local statistics transmitted by the \glspl{ap} \gls{ota}. We provided expressions for the \gls{mse} of these sufficient statistics and the \gls{se} of the \glspl{ue}, and  conducted a comprehensive performance study to highlight the benefits of the \gls{ota} framework compared to  conventional digital fronthaul. 

We showed that a wireless fronthaul with an \gls{ota} computation framework resulted in a significant reduction in the fronthaul load.
In our simulations,   a digital fronthaul requires more than $8$ bits per real symbol to achieve  reasonably good performance, and at least $16$ bits to match the performance of the analog \gls{ota} scheme. Therefore, the \gls{ota}  framework is a promising alternative to  conventional wired fronthaul. The  concept of \gls{ota}  wireless fronthaul signaling also brings several new, interesting research challenges and opportunities that remain for exploration in 
 future work.}

\bibliographystyle{IEEEtran}
\bibliography{IEEEabrv,reff}
\end{document}

%% file: header.tex
\documentclass[journal, final, twocolumn, 10pt]{IEEEtran}
\usepackage[utf8]{inputenc}
\usepackage{amsmath}
\usepackage{amssymb}
\usepackage{url}
\usepackage{graphicx}
\usepackage{epstopdf}
\usepackage{caption}
\usepackage{comment}
\usepackage{subcaption}
\usepackage{cancel}
\usepackage{soul}
\usepackage{multirow}
\usepackage{amssymb}
\usepackage{setspace}
\usepackage[noadjust]{cite}
\usepackage{xspace}
\usepackage{algorithm}
\usepackage{tabularx}
\usepackage{changepage}

\usepackage{graphicx,epsfig,bm}
\usepackage{float}
\usepackage{cite}
\usepackage{amsmath,epsfig} 
\usepackage{times}
\usepackage{enumerate,type1cm}
\usepackage{amsfonts,relsize}
\usepackage{bm}
\usepackage{amssymb}
\usepackage{relsize}
\usepackage{fancybox}
\usepackage{algorithmic}
\usepackage{amssymb}
\usepackage{graphicx,epsfig,bm}
\usepackage{amsmath,epsfig} 
\usepackage{times}
\usepackage{enumerate,type1cm}
\usepackage{amsfonts,relsize}
\usepackage{bm}
\usepackage{amssymb}
\usepackage{relsize}
\usepackage{fancybox}
\usepackage{bbm}
\usepackage{tikz}
\usetikzlibrary{fit,positioning}
\usetikzlibrary{bayesnet}
\usetikzlibrary{shapes,decorations}
\usepackage[english]{babel}
\usepackage{xpatch}
\usepackage{mathabx}

\usepackage{pgf}
\usepackage{pgfplots}
\usepackage{tikz}
\usetikzlibrary{spy}
\pgfplotsset{compat=1.17}

\usepackage{subcaption}
\usepackage{eqparbox}
\usepackage{makecell}
\usepackage{svg}

\usepackage[xindy]{glossaries}
\usepackage{xcolor}
\usepackage{url}
\def\BibTeX{{\rm B\kern-.05em{\sc i\kern-.025em b}\kern-.08em
		T\kern-.1667em\lower.7ex\hbox{E}\kern-.125emX}}

\newcommand{\numAPs}{L}
\newcommand{\numAntennasPerAP}{N}
\newcommand{\numUEs}{K}
\newcommand{\numAntennasCPU}{M}

\newcommand{\tauu}{\tau_{\rm u}}

\newcommand{\chan}[1]{\mathbf{h}_{{#1}}}

\newcommand{\chanMat}[1]{\mathbf{H}_{{#1}}}

\newcommand{\chanMatCent}[1]{\mathbf{G}_{{#1}}}

\newcommand{\norm}[1]{\left\Vert#1\right\Vert}
\newcommand{\absL}[1]{\left\vert#1\right\vert}
\newcommand{\abs}[1]{\vert#1\vert}

\newcommand{\complexm}[2]{\mathbb{C}^{#1\times #2}}

\newcommand{\gramian}[1]{#1^H#1}

\newcommand{\CN}[2]{\mathcal{CN}{(#1,#2)}}
\newcommand{\expect}[1]{\mathbb{E}\{#1\}}
\newcommand{\expectL}[1]{\mathbb{E}\left\{#1\right\}}
\newcommand{\trace}[1]{{\rm tr}\left(#1\right)}

\newcommand{\bluez}[1]{{\color{black} #1}}

\newcommand{\colr}[1]{{\color{black} #1}}

\loadglsentries{abbrev}
\usepackage{ntheorem}

\long\def\symbolfootnote[#1]#2{\begingroup%
	\def\thefootnote{\fnsymbol{footnote}}\footnote[#1]{#2}\endgroup}

\newcommand{\beq}{\begin{equation}}
\newcommand{\eeq}{\end{equation}}
\newcommand{\beqa}{\begin{eqnarray}}
\newcommand{\eeqa}{\end{eqnarray}}

\usepackage{xcolor}
\definecolor{forestgreen}{rgb}{0.133, 0.545, 0.133}

\usepackage{float}
\usepackage{colortbl}

\DeclareMathOperator*{\argmin}{arg\,min}
\DeclareMathOperator*{\argmax}{arg\,max}

\tikzset{
	startstop/.style={
		rectangle, 
		rounded corners,
		minimum width=3cm, 
		minimum height=0.5cm,
		align=center, 
		draw=black, 
	},
	process/.style={
		rectangle, 
		minimum width=3cm, 
		minimum height=0.5cm, 
		align=center, 
		draw=black, 
	},
	decision/.style={
		rectangle, 
		minimum width=3cm, 
		minimum height=0.5cm, align=center, 
		draw=black, 
	},
	arrow/.style={thick,->,>=stealth},
	dec/.style={
		ellipse, 
		align=center, 
		draw=black, 
	},
}


\makeatletter
\def\BState{\State\hskip-\ALG@thistlm}

\DeclareCaptionLabelSeparator{periodspace}{.\quad}
	{\end{adjustwidth}}

\newcommand{\gram}{A}
\newcommand{\mf}{t}

\usepackage{stfloats}

\newcommand{\ceil}[1]{\left\lceil #1 \right\rceil}

\makeatletter
\newcommand{\removelatexerror}{\let\@latex@error\@gobble}
\makeatother


%% file: abbrev.tex
%
%
\newacronym{ai}{AI}{artificial intelligence}
\newacronym{aoa}{AOA}{angle-of-arrival}
\newacronym{aod}{AOD}{angle-of-departure}
\newacronym{ap}{AP}{access point}
\newacronym{asic}{ASIC}{application specific integrated circuit}
\newacronym{awgn}{AWGN}{additive white Gaussian noise}
\newacronym{blu}{BLU}{best linear unbiased}
\newacronym{blue}{BLUE}{best linear unbiased estimator}
\newacronym{ber}{BER}{bit error rate}
\newacronym{ce}{CE}{channel estimation}
\newacronym{cw}{CW}{continuous wave}
\newacronym{cdf}{CDF}{cumulative distribution function}
\newacronym{ced}{CED}{cumulative energy density}
\newacronym{clk}{CLK}{clock}
\newacronym{csi}{CSI}{channel state information}
\newacronym{csp}{CSP}{contact service point}
\newacronym{cpu}{CPU}{central processing unit}
\newacronym{crlb}{CRLB}{Cram\'er-Rao lower bound}
\newacronym{cu}{CU}{central unit}
\newacronym{dc}{DC}{direct current}
\newacronym{dl}{DL}{downlink}
\newacronym{ds}{ODS}{digital scheme}
\newacronym{dlt}{DLT}{Distributed Ledger Technology}
\newacronym{dp}{DP}{differencial privacy}
\newacronym{dram}{DRAM}{dynamic random-access memory}
\newacronym{du}{DU}{distributed unit}
\newacronym{ecc}{ECC}{elliptic curve cryptography}
\newacronym{ecdlp}{ECDLP}{elliptic curve discrete logarithm problem}
\newacronym{ecsp}{ECSP}{edge computing service point}
\newacronym{eirp}{EIRP}{equivalent isotropically radiated power}
\newacronym{em}{EM}{electromagnetic}
\newacronym{en}{EN}{energy neutral}
\newacronym{e2e}{E2E}{End-to-End}
\newacronym{fa}{FA}{federation anchor}
\newacronym{fdd}{FDD}{frequency division duplexing}
\newacronym{fh}{FH}{fronthaul}
\newacronym{fhss}{FHSS}{frequency hopping spread spectrum}
\newacronym{fl}{FL}{federated learning}
\newacronym{gdpr}{GDPR}{general data protection regulation}
\newacronym{ia}{IA}{initial access}
\newacronym{ic}{IC}{integrated circuit}
\newacronym{ioe}{IoE}{internet of everything}
\newacronym{iot}{IoT}{internet of things}
\newacronym{id}{ID}{information decoding}
\newacronym{kpi}{KPI}{key performance indicator}
\newacronym{lca}{LCA}{life cycle assessment}
\newacronym{ls}{LS}{least squares}
\newacronym{lis}{LIS}{large intelligent surface}
\newacronym{liion}{Li-ion}{lithium-ion}
\newacronym{llr}{LLR}{log-likelihood ratio}
\newacronym{los}{LoS}{line-of-sight}
\newacronym{lmo}{LMO}{lithium ion manganese oxide}
\newacronym{lte}{LTE}{long-term evolution}
\newacronym{map}{MAP}{maximum a posteriori}
\newacronym{mf}{MF}{matched filter}
\newacronym{ml}{ML}{maximum-likelihood}
\newacronym{mle}{MLE}{maximum likelihood estimator}
\newacronym{mvu}{MVU}{minimum variance unbiased}
\newacronym{mse}{MSE}{mean-square error}
\newacronym{mmse}{MMSE}{minimum mean square error}
\newacronym{lmmse}{LMMSE}{linear minimum mean-square error}
\newacronym{ldpc}{LDPC}{low-density parity-check}
\newacronym{mmwave}{mmWave}{millimeter wave}
\newacronym{mimo}{MIMO}{multiple-input multiple-output}
\newacronym{miso}{MISO}{multiple-input single-output}
\newacronym{mpc}{MPC}{multipath component}
\newacronym{mrc}{MRC}{maximum ratio combining}
\newacronym{mrt}{MRT}{maximum ratio transmission}
\newacronym{nmse}{NMSE}{normalized mean square error}
\newacronym{nods}{NODS}{non-orthogonal digital scheme}
\newacronym{ofdm}{OFDM}{orthogonal frequency-division multiplexing}
\newacronym{ota}{OTA}{over-the-air}
\newacronym{ods}{ODS}{orthogonal digital scheme}
\newacronym{nb}{NB}{narrowband}
\newacronym{nr}{NR}{New Radio}
\newacronym{pa}{PA}{power amplifier}
\newacronym{pdp}{PDP}{power delay profile}
\newacronym{per}{PER}{packet error rate}
\newacronym{pet}{PET}{privacy enhancing technology}
\newacronym{pki}{PKI}{public key infrastucture}
\newacronym{pl}{PL}{path loss}
\newacronym{pc}{PC}{pilot count}
\newacronym{qrrls}{QR-RLS}{QR decomposition based recursive least squares}
\newacronym{qam}{QAM}{quadrature amplitude modulation}
\newacronym{qos}{QoS}{quality-of-service}
\newacronym{re}{RE}{radio element}
\newacronym{rf}{RF}{radio frequency}
\newacronym{rfid}{RFID}{radio frequency identification}
\newacronym{ris}{RIS}{reflective intelligent surface}
\newacronym{rms}{RMS}{root mean square}
\newacronym{rls}{RLS}{recursive least squares}
\newacronym{rss}{RSS}{received signal strength}
\newacronym{rssi}{RSSI}{received signal strength indicator}
\newacronym{rw}{RW}{RadioWeaves}
\newacronym{sa}{SA}{synchronization anchor}
\newacronym{sdg}{SDG}{Sustainable Development Goal}
\newacronym{sdn}{SDN}{software-defined network}
\newacronym{se}{SE}{spectral efficiency}
\newacronym{ser}{SER}{symbol error rate}
\newacronym{sinr}{SINR}{signal-to-interference-plus-noise ratio}
\newacronym{siso}{SISO}{single-input single-output}
\newacronym{slam}{SLAM}{simultaneous localization and mapping}
\newacronym{snr}{SNR}{signal-to-noise ratio}
\newacronym{srls}{SRLS}{standard recursive least squares}
\newacronym{svd}{SVD}{singular value decomposition}
\newacronym{sp}{SP}{service point}
\newacronym{sram}{SRAM}{static random-access memory}
\newacronym{tdd}{TDD}{time division duplexing}
\newacronym{tdoa}{TDOA}{time-difference-of-arrival}
\newacronym{toa}{TOA}{time-of-arrival}
\newacronym{tx}{TX}{transmitter}
\newacronym{rx}{RX}{receiver}
\newacronym[longplural={user equipment}]{ue}{UE}{user-equipment}
\newacronym{uatf}{UatF}{use-and-then-forget}
\newacronym{uhf}{UHF}{ultra high frequency}
\newacronym{ula}{ULA}{uniform linear array}
\newacronym{upa}{UPA}{uniform planar array}
\newacronym{ul}{UL}{uplink}
\newacronym{ura}{URA}{uniform rectangular array}
\newacronym{uwb}{UWB}{ultrawideband}
\newacronym{wb}{WB}{wideband}
\newacronym{wpt}{WPT}{wireless power transfer}
\newacronym{zf}{ZF}{zero-forcing}

\newacronym{lpt}{LPT}{laser power transfer}
\newacronym{rfpt}{RFPT}{radio frequency power transfer}
\newacronym{ipt}{IPT}{inductive power transfer}
\newacronym{cpt}{CPT}{capacitive power transfer}
\newacronym{apt}{APT}{acoustic power transfer}
\newacronym{cfo}{CFO}{carrier frequency offset}
\newacronym{lo}{LO}{local oscillator}
\newacronym{if}{IF}{intermediate frequency}
\newacronym{duc}{DUC}{digital up conversion}
\newacronym{ddc}{DDC}{digital down conversion}
\newacronym{dsp}{DSP}{digital signal processing}
\newacronym{vco}{VCO}{voltage-controlled oscillator}
\newacronym{pll}{PLL}{phase-locked loop}
\newacronym{lna}{LNA}{low noise amplifier}
\newacronym{dac}{DAC}{digital-to-analog converter}
\newacronym{adc}{ADC}{analog-to-digital converter}
\newacronym{de}{DE}{drain efficiency}
\newacronym{pae}{PAE}{power-added efficiency}
\newacronym{sar}{SAR}{specific absorption rate}
\newacronym{mppt}{MPPT}{maximum power point tracking}
\newacronym{isi}{ISI}{intersymbol interference}
\newacronym{pps}{PPS}{pulse per second}
\newacronym{icnirp}{ICNIRP}{International Commission on Non-Ionizing Radiation Protection}
\newacronym{fd}{FD}{Front-haul distance}
\newacronym{wd}{WD}{Wireless distance}
\newacronym{ntp}{NTP}{Network Time Protocol}
\newacronym{ptp}{PTP}{Precision Time Protocol}
\newacronym{wr}{WR}{White Rabbit}
\newacronym{wlan}{WLAN}{wireless local-area network}